\documentclass[%
superscriptaddress,
longbibliography,
preprint,
amsmath,amssymb,
aps,
prx,
]{revtex4-1}

\usepackage{ulem}
\usepackage{amsmath}
\usepackage{amsfonts}
\usepackage{graphicx}
\usepackage{dcolumn}
\usepackage{bm}
\usepackage{lipsum}
\usepackage{hyperref}
\usepackage{multirow}

\newcommand{\env}{\rho\textbf{}}
\newcommand{\covariance}{\Sigma}
\newcommand{\Ntrain}{N_\text{T}}
\newcommand{\Nsubset}{{n \choose 2}}

\begin{document}

\title{Bayesian Force Fields from Active Learning for Simulation of Inter-Dimensional Transformation of Stanene}

\author{Yu Xie}
\email{xiey@g.harvard.edu}
\affiliation{John A. Paulson School of Engineering and Applied Sciences, Harvard University, Cambridge, MA 02138, USA}

\author{Jonathan Vandermause}
\affiliation{John A. Paulson School of Engineering and Applied Sciences, Harvard University, Cambridge, MA 02138, USA}

\author{Lixin Sun}
\affiliation{John A. Paulson School of Engineering and Applied Sciences, Harvard University, Cambridge, MA 02138, USA}

\author{Andrea Cepellotti}
\affiliation{John A. Paulson School of Engineering and Applied Sciences, Harvard University, Cambridge, MA 02138, USA}

\author{Boris Kozinsky}
\email{bkoz@seas.harvard.edu}
\affiliation{John A. Paulson School of Engineering and Applied Sciences, Harvard University, Cambridge, MA 02138, USA}
\affiliation{%
Robert Bosch LLC, Research and Technology Center,
Cambridge, Massachusetts 02142, USA}



\begin{abstract}

We present a way to dramatically accelerate Gaussian process models for interatomic force fields based on many-body kernels by mapping both forces and uncertainties onto functions of low-dimensional features. This allows for automated active learning of models combining near-quantum accuracy, built-in uncertainty, and constant cost of evaluation that is comparable to classical analytical models, capable of simulating millions of atoms. Using this approach, we perform large scale molecular dynamics simulations of the stability of the stanene monolayer. We discover an unusual phase transformation mechanism of 2D stanene, where ripples lead to nucleation of bilayer defects, densification into a disordered multilayer structure, followed by formation of bulk liquid at high temperature or nucleation and growth of the 3D bcc crystal at low temperature. The presented method opens possibilities for rapid development of fast accurate uncertainty-aware models for simulating long-time large-scale dynamics of complex materials.  
\end{abstract}

\maketitle


\section*{\label{sec:intro}Introduction}

Density functional theory (DFT) 
is one of the most successful methods for simulating condensed matter thanks to a reasonable accuracy for a wide range of systems. Ab-initio molecular dynamics (AIMD) offers a way to simulate the atomic motion using forces computed at the DFT level.
Unfortunately, computational requirements limit the time-scale and size of AIMD simulations to a few hundred atoms for a few hundred picoseconds of time, precluding investigation of phase transitions and heterogeneous reactions.
Such large-scale molecular dynamics (MD) simulation must resort to empirically derived analytical interatomic force fields with fixed functional form \cite{van2001reaxff, chenoweth2008reaxff, baskes1987application, lindsay2010optimized}, trading accuracy and transferability for larger length and time scales.
Classical analytical force fields often do not match the accuracy of ab-initio results, limiting simulations to describing results qualitatively at best or, at worst, deviating from the correct behavior.
In order to broaden the reach of computational simulations, it would be desirable to compute forces with ab-initio accuracy at the same cost of classical interatomic force fields.

In recent years, machine learning (ML) algorithms emerged as powerful tools in regression and classification problems.
This interest has inspired several works to develop ML algorithms for interatomic force fields, for example 
neural-networks (NN) \cite{behler2007generalized, behler2011atom, behler2011neural, behler2015constructing, behler2016perspective, mailoa2019fast, smith2017ani, batzner2021se}, 
MTP \cite{shapeev2016moment, podryabinkin2017active, hodapp2020operando}, FLARE \cite{vandermause2020fly, Lim2020},
GAP \cite{bartok2010Gaussian, bartok2013representing, bartok2015g, bartok2018machine}, 
SNAP \cite{osti_1183138}, 
SchNet \cite{schutt2017schnet, schutt2018schnet, schutt2018schnetpack}, 
DeePMD \cite{zhang2018deep, wang2018deepmd, han2017deep}, among others. 
All these machine learning potentials can make predictions at near ab-initio accuracy, while greatly reducing the computational cost compared to DFT. 

However, most machine learning models have no predictive distributions which provide uncertainty for energy/force predictions. Without uncertainty, the training data is generally selected from ab-initio calculations via a manual or random scheme. Determining the reliability of the force fields then becomes difficult, which could result in untrustworthy configurations in the molecular dynamics simulations.
Therefore, uncertainty quantification is a highly desirable capability \cite{musil2019fast,peterson2017addressing}.
Neural network potentials, e.g. ANI \cite{smith2017ani, smith2018less} uses ensemble disagreement as an uncertainty measure. This statistical approach is, however, not guaranteed to yield reliable calibrated uncertainty. In addition, NN approaches usually require tens of thousands of data for training, and are a few orders of magnitude slower than analytical force fields.
Bayesian models are promising for uncertainty quantification in atomistic simulations since they have an internal principled uncertainty quantification mechanism, the variance of the posterior prediction distribution, which can be used to keep track of the error on forces during a molecular dynamics (MD) run.
For instance, Jinnouchi \textit{et al.}\ \cite{jinnouchi2019phase, jinnouchi2019fly} used high-dimensional SOAP descriptors with Bayesian linear regression \cite{bartok2013representing}.
Gaussian Process (GP) regression \cite{glielmo2018efficient, glielmo2017accurate, glielmo2020building,bartok2010Gaussian,bartok2013representing} is a Bayesian method that has been shown to learn accurate forces with relatively small training data sets.
Bartok \textit{et al.}\ \cite{bartok2018machine} used GP uncertainty with the GAP/SOAP framework to obtain only qualitative estimates of the force field’s accuracy. 
Recently Vandermause \textit{et al.}\ \cite{vandermause2020fly} demonstrated a GP-based Bayesian active learning (BAL) scheme in the FLARE framework, utilizing rigorously principled and quantitatively calibrated uncertainty, applying it to a variety of multi-component materials.

In the most common form, however, GP models require using the whole training data set for prediction of both the force and uncertainty, meaning that the computational cost of prediction grows linearly with the size of the training set, and so accuracy increases together with computational cost.
In complex multi-component structures, more than $O(10^2)$ training structures or $O(10^3)$ local environments are typically required to construct an accurate GP model, which makes predictions slow \cite{chmiela2018towards, Zuo2020}.
Because of the linear scaling of GP, the on-the-fly training becomes slower as more data are collected. This precludes the active learning of GP on larger system sizes which may be needed due to finite size effects, as well as longer time scales needed to explore phase space more thoroughly.
To accelerate the on-the-fly training, fast and lossless mappings as approximations of both the force and uncertainty are desired, such that the mappings can replace GP to make predictions during the Bayesian active learning.

The force mapping has been addressed by Glielmo \textit{et al.}\cite{glielmo2018efficient, glielmo2017accurate, glielmo2020building, glielmo2018efficient} and Vandermause \textit{et al.} \cite{vandermause2020fly}. 
They noted that, for a suitable choice of the $n$-body kernel function, it is possible to decompose GP force/energy prediction into a summation of low-dimensional functions of atomic positions.
As a result, one can construct a parametric mapping of the $n$-body kernel function combined with a fixed training set. This approach reaches a constant scaling, which increases the speed of the GP model without accuracy loss.
We also present the formalism of the force mapping in the \textit{Methods} section.

Since the uncertainty plays a central role in the on-the-fly training \cite{vandermause2020fly} and also suffers from the linear-scaling complexity with respect to the training set size, it is desirable to have a similar mapping of the uncertainty as that of the force. However, this was not attempted in \cite{vandermause2020fly, glielmo2018efficient, glielmo2017accurate, glielmo2020building}, since the mathematical form of the predictive variance results in decomposition with twice the dimension of the corresponding mean function, dramatically increasing the computational cost of their evaluation with spline interpolation. 
Thus, to date there is no method that combines high accuracy, modest training data requirement and fast prediction of both forces and their principled Bayesian uncertainty.
In this work, we introduce a dimensionality reduction technique for the uncertainty mapping, such that the interpolation can be done on the same dimensionality as the force mapping, enabling development and application of efficient \textit{Bayesian force field} (BFF) models for complex materials.
The mathematical formalism of the uncertainty mapping is presented in the \textit{Methods} section.

In this article, we present an accurate mapping of both force (mean) and uncertainty (variance) implemented as the \textit{Mapped Gaussian Process} (MGP) method. As a result, the MGP method benefits from the capability of quantifying uncertainty, while at the same time retaining cost independent of training size.
The original BAL with GP can then be accelerated by the fast evaluation of forces and uncertainties of MGP. The training can also be extended to larger system sizes and longer time scales that are challenging for the full GP.
To illustrate the performance for large-scale dynamics simulations, we incorporate the MGP force field with mean-only mapping in the parallel MD simulation code LAMMPS \cite{plimpton1995fast}, and apply it to the investigation of phase transformation dynamics. The MGP force field is shown to be efficient for large-scale (million atom) simulations, achieving speeds comparable to classical analytical force fields, several orders of magnitude faster than available NN or full GP approaches.

As a test application, we focus on stanene, a 2D material that has recently gathered attention as a quantum spin Hall insulator \cite{tang2014stable}. Moreover, the only published force field for stanene is, to the best of our knowledge, the bond-order potential by Cherukara \textit{et al.}\cite{cherukara2016ab}, which is fitted to capture stanene's low-temperature crystalline characteristics but, as common to many empirical interatomic potentials, suffers in accuracy near the melting temperature. 
In particular, we show that MGP is capable of rapidly learning to describe a wide range of temperatures, below and around the melting temperature, and that we can efficiently monitor the uncertainty on forces at each time step of the molecular dynamics, and use this capability to iteratively increase the accuracy of the force field by hierarchical active training. Using parallel simulations of large-scale structures, we characterize the unusual phase transition, where the 2D monolayer transforms to bcc bulk Sn at the temperature of 200K, and melts to ultimately form a 3D liquid phase at the temperature of 500K.


\section*{\label{sec:simu}Results}


\subsection*{Accelerated Bayesian Active Learning with Mapped Gaussian Processes\label{subsec:bal}}

In a MD simulation, it is likely that the system will evolve to atomic configurations unseen before, and are far from those in the training set.
In this situation, the uncertainties of the predictions will grow to large values which may be considered unsatisfactory.
Therefore, it is desirable to obtain an accurate estimate of the forces for such configurations using a relatively expensive first principles calculation, and add this new information to the training set of the GP regression model.

This procedure is referred to as \textit{Bayesian Active Learning} (BAL), where training examples are added on the fly as more information is obtained about the problem.
Bayesian machine learning models such as GP are particularly well-suited to such uncertainty-based active learning approaches, as they provide a well-defined probability distribution over model outputs which can be used to extract rigorous estimates of model uncertainty for any test point \cite{bishop2006pattern, murphy2012machine}. 

Here we adopt BAL to achieve automatic training of models for atomic forces, expanding on our earlier workflow \cite{vandermause2020fly}.
This way the accuracy of the GP model increases with time, particularly in the configuration regions that are less explored and likely to be inaccurate.

\begin{figure*}[h!]
    \centering
    \includegraphics{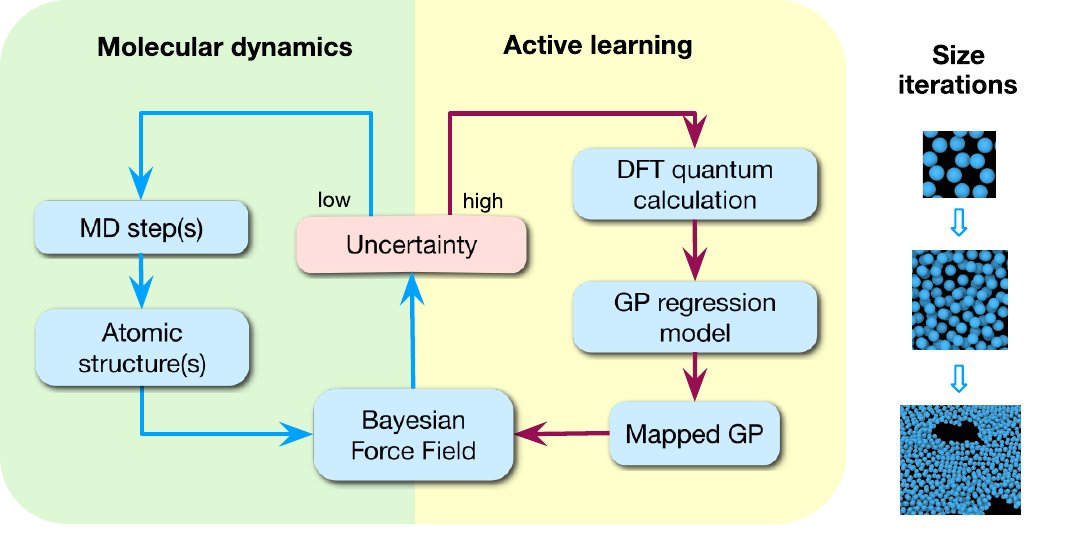}
    \caption{Bayesian active learning (BAL) workflow with MGP. Iterations of BAL can be run for different system sizes to refine the force field.}
    \label{fig:flow}
\end{figure*}

As mentioned above, GP regression cost scales linearly with the training set size, so the prediction of forces and uncertainties becomes more expensive as the BAL algorithm keeps adding data to the training set, hindering simulations of complex materials systems. 
The MGP approach is an essentially exact mapping of both mean and variance of the predicted forces from a full GP model for a fixed training set to a low-dimensional spline model. 
This approach has the ability to accelerate prediction of not only the mean, but also GP's internal uncertainty without loss of accuracy.
MGP is incorporated into the BAL workflow as depicted in Fig.\ref{fig:flow}. In this scheme, the mapping is constructed every time the GP model is updated. Atomic forces and their uncertainties for MD are produced by MGP, reducing the computational cost of the full original GP prediction by orders of magnitude.
We stress that not only is the MGP BFF faster than the original GP, but also its speedup is more pronounced as the training data set size $\Ntrain$ increases. This is especially critical in complex systems with multiple atomic and molecular species and phases, where more training configurations are needed.

In addition, fast evaluation of forces and their variances enables training the force field model on larger system sizes. 
Therefore, we can train the BFF using a hierarchical scheme, i.e. for different system sizes, we can run several iterations of active learning. First, a small system size is used to train the BFF using the BAL workflow. 
Then, we perform BAL on a larger system size, where DFT calculations are more expensive but are needed less frequently, in order to capture potentially new unexplored configurations such as defects that are automatically identified as high uncertainty by the model of the previous step. 
We note that DFT calculations scale as the 3rd power of the number of electrons in the system, so it is efficient to perform most of the training at smaller sizes, and to only require re-training on fewer large-scale structures. This way the hierarchical training scheme helps to overcome finite size effects and explore phenomena that cannot be captured with small system sizes, such as phase transformations.


\subsection*{Case Study: 2D to 3D Transformation of Stanene\label{subsec:case}}

To illustrate our method, we apply the mapped on-the-fly workflow (Fig.\ref{fig:flow}) to study the phase transformation of stanene, a two-dimensional slightly buckled honeycomb lattice of tin. This structure has received much attention recently due to its unique topological electronic properties. Stanene can be synthesized by molecular beam epitaxy \cite{zhu2015epitaxial} and has been shown to have topologically protected states in the band gap \cite{deng2018epitaxial}. Moreover, tri-layer stanene is a superconductor \cite{liao2018superconductivity}. 
As a demonstration of our methodology, here we want to focus on the less studied aspects of thermodynamic stability and phase transition mechanisms of free standing stanene. 
We note that a substrate is currently always used in the growth of stanene monolayers, and the interaction between the monolayer and the substrate will be investigated in a future work. By studying the intrinsic thermodynamics properties of the free-standing stanene, we are able to examine its stability and decomposition timescales as a function of temperature, which are relevant for device engineering.

Despite the intense interest, capability is still lacking for accurate modeling of thermodynamic properties and phase stability of stanene, and most other 2D materials,  for which $\sim$nm scale supercells are required. 
Such simulations are beyond the ability of DFT, and to the best of our knowledge, the only published classical force field for stanene is a bond-order potential (BOP) \cite{cherukara2016ab}. 
Our focus is on developing an accurate potential to model the 2D melting process, which is a notoriously difficult problem to tackle with existing classical force fields. 
In fact, describing the melting mechanism of a 2D material is a highly non-trivial problem, since it can be characterized by specific behaviors such as defect formation and nucleation. 
For example, the melting of graphene is thought to occur in a temperature range between 4000K and 6000K \cite{ganz2017initial, PhysRevB.91.045415, zakharchenko2011melting}, and seems to be characterized by the formation of defects, while above 6000K the structure melts directly into a totally disordered configuration. 
Since bulk tin melts at a much lower temperature ($\sim 500$K) than graphite ($\sim 4300$K), we expect that stanene will transform at a relatively low temperature. However, little is known about its melting mechanism and transition temperature. 
The MGP method is uniquely suited to describe the transformation processes, as it is scalable to large systems and capable of accurately describing any configuration that is sufficiently close to the training data.


To train the force field for stanene, we start by training the MGP BFF using the BAL loop on a small system size, and gradually increase the size of the simulation, iteratively improving upon the force field developed for the smaller size.

The DFT settings and the hyperparameters of GP are shown in Supplementary Table 1. The parameters of MGP model are shown in Supplementary Table 2. The atomic configurations are plotted by OVITO \cite{ovito}.

\begin{figure*}[h!]
    \centering
    \includegraphics{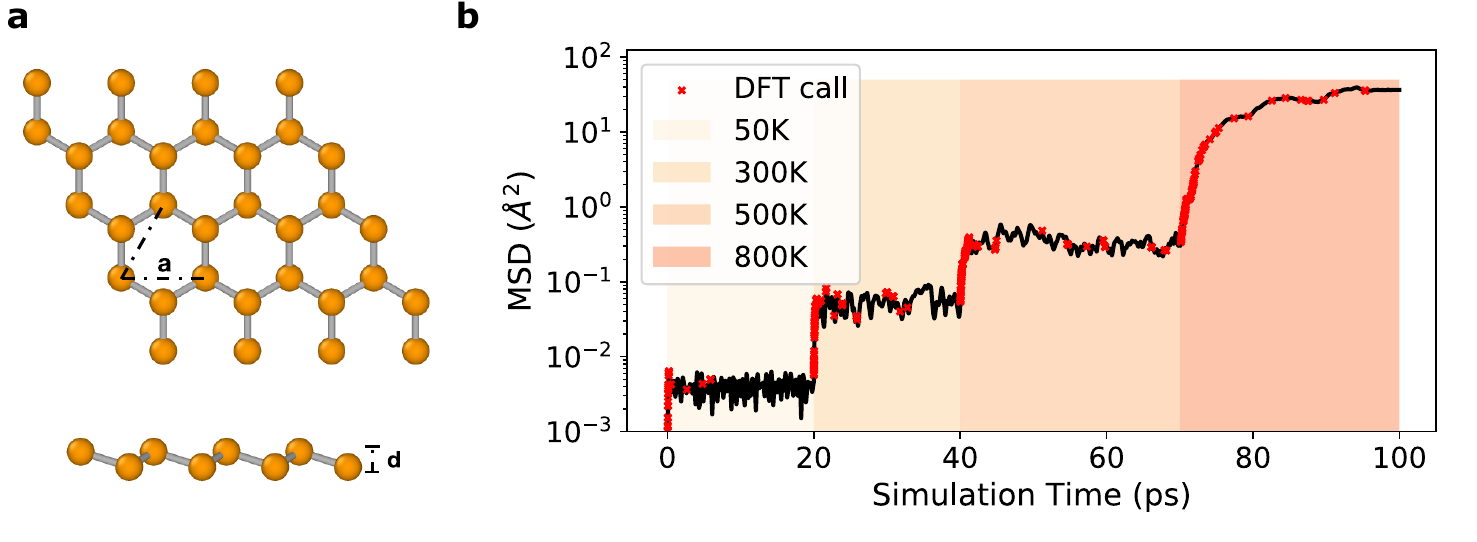}
    \caption{Bayesian active learning of stanene.
    \textbf{a.} Pristine stanene, top view and front view. 
    \textbf{b.} Atomic mean square displacement (MSD) of the on-the-fly training trajectory. The solid black line shows the MSD of the 32 atoms, the color blocks indicate time intervals corresponding to different temperatures, and the red dots represent events when DFT calculations are called as the uncertainty predictions exceed the data acquisition decision threshold and new data is added to the training set.}
    \label{fig:cell_msd}
\end{figure*}

The first iteration of training is started with  a $4\times 4\times 1$ supercell of stanene (32 atoms in total) and a 20 ps long MD simulation at a temperature of 50K. Subsequently, we increase the temperature of the system at 300K, 500K and 800K by velocity rescaling, and let the system evolve for 20, 30 and 30 ps respectively. This way we efficiently augment the training set with relevant configurations.
During the simulation, DFT was called whenever the uncertainty on any single force component exceeded the current optimized noise parameter of the Gaussian process, after which the $N=1$ atomic environment with the highest force uncertainty was added to the training set.
Fig.\ref{fig:cell_msd}b shows the mean square displacement (MSD) of atoms during the on-the-fly BAL training. As one would expect, the DFT is called multiple times at the beginning of the MD simulation, since the model needs to build an initial training data set to make force predictions. Additionally, the DFT is called whenever temperature is increased, as the trajectory explores new configurations that have not been visited before.
Clearly, the variance increases rapidly when temperature is increased. 

In this small simulation, the stanene monolayer decomposes at 800K, where Sn atoms start moving around the simulation cell. This allows us to augment the training set with such disordered configurations. However, the small size of the simulation cell does not yet allow us to draw conclusions about the melting process of stanene: for this, we need a larger simulation cell.

\begin{figure*}[h!]
    \includegraphics{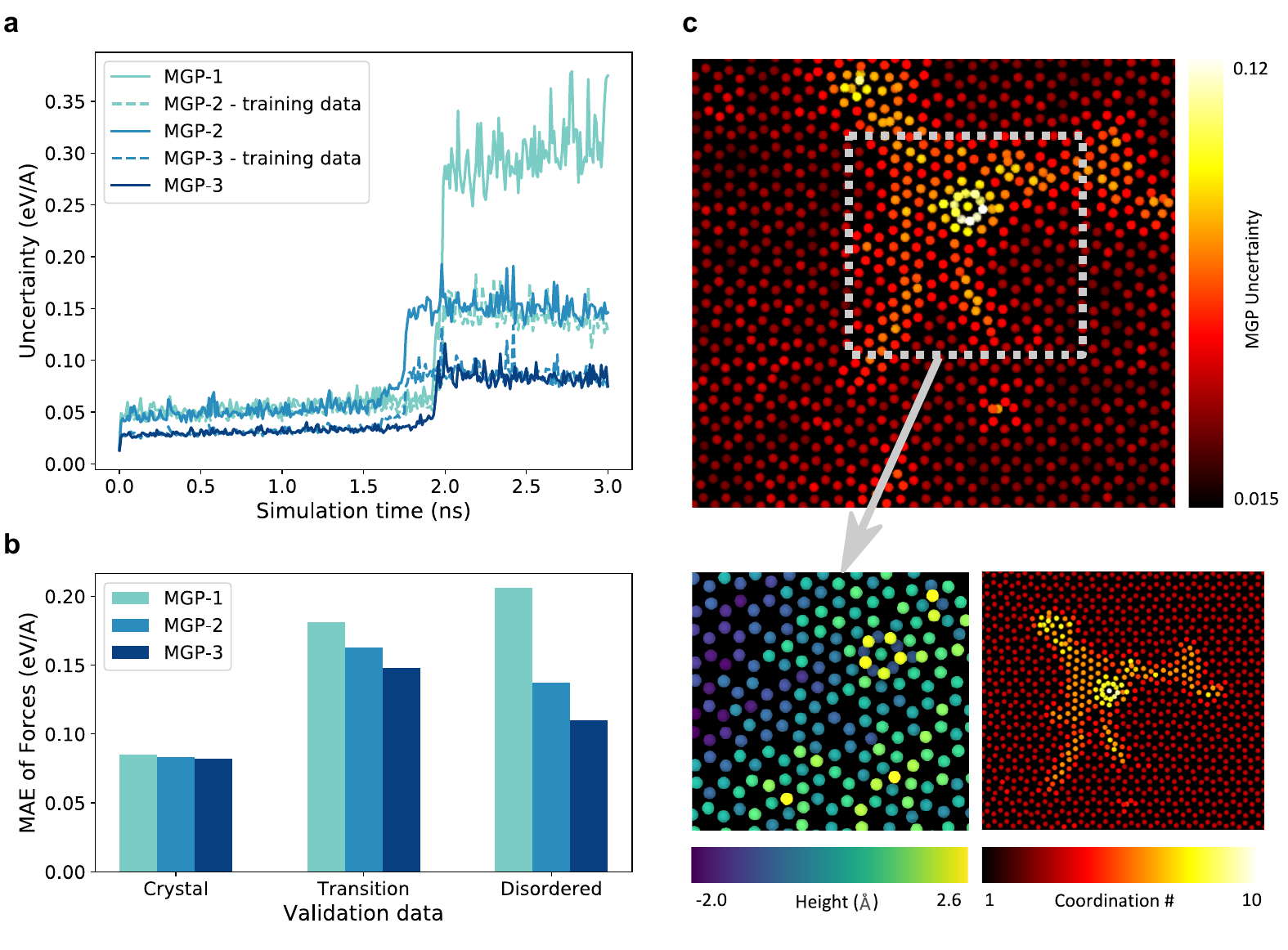}
    \caption{Uncertainty and accuracy in the Bayesian active learning.
    \textbf{a.} Uncertainty of the force predictions in the whole MD trajectory decreases in three iterations.
    \textbf{b.} Mean absolute errors (MAE) of MGP forces against DFT in three iterations, validated on atomic structures of different phases.
    \textbf{c.} Upper: A snapshot of 2048 atoms from MD simulation at 500K, colored by MGP uncertainty. The outlined area has highest predicted uncertainty, since it contains a defect and a compact region not known to the first MGP model. Lower-left: Zoomed-in structure of the region of high uncertainty, colored by atomic heights (z-coordinate) to show the bilayer structure of the defect. Lower-right: The same snapshot colored by coordination number, which shows strong correlation with the uncertainty measure. }
    \label{fig:uncertainty_mgp-abc}
\end{figure*}


We continue with our hierarchical active learning scheme to refine the force field, but retaining the training data from the previous smaller MD run. We now increase the number of atoms in the simulation, which allows us to explore additional atomic configurations and improve the quality of the model. 
Since the system becomes larger and more computationally expensive, instead of updating the model at each learning step, we update it at the end of each MD simulation. This ``batching" approach is one of many possible training schemes and is chosen for reasons of efficiency.  The procedure is as below:

\begin{enumerate}
    \item Run MD simulation on a system of a given size with MGP;
    \item Predict mapped uncertainties for the frames from the MD trajectory;
    \item Select frames of highest uncertainties in different structures (crystal, transition state and disordered phase) for additional DFT calculations;
    \item Add atomic environments from those frames based on uncertainty to the training set of GP;
    \item Construct a new MGP BFF using the updated GP;
	\item Repeat the process for same or larger simulation using the new BFF.
\end{enumerate}

The above steps form one iteration, and can be run several times to refine the force field. In our force field training for stanene, we performed one iteration of 32 atoms as described before, and two iterations all on the system size of 200 atoms at 500K, a relatively high temperature to explore diverse structures in the configuration phase space.
We denote the MGP BFF from the on-the-fly training of 32 atoms as MGP-1, the ones from the two iterations of training on 200 atoms as MGP-2 and MGP-3.
MD are simulated by MGP-1, 2, 3 and the predicted uncertainties of the frames from the three MD trajectories predicted by MGP BFF are shown in Fig.\ref{fig:uncertainty_mgp-abc}a. 
First, there is a drastic increase of uncertainty within each trajectory at a certain time. This increase corresponds to the transition from crystal lattice to a disordered phase, so the uncertainty acts as an automatic indicator of the structural change. 
Next, the uncertainties of MGP-1, 2, 3 decrease iteration by iteration, indicating our model becomes more confident as it collects more data. The mean absolute error (MAE) of force predictions of MGP BFF against DFT is investigated, as presented in Fig.\ref{fig:uncertainty_mgp-abc}b, showing that MGP BFF becomes also more accurate with each subsequent training iterations.
The accuracy of force prediction from MGP force field is also much higher than BOP \cite{cherukara2016ab} for different phases of stanene, as shown in Supplementary Table 4.

It is worth noting that in the transition process of the simulation by MGP-1, there appear a bi-layer stacked defect and a specific compact triangular lattice around the defect. It is a visibly new ordered structure different from the hexagonal stanene, and is thus missing in the initial training set, as shown in Fig.\ref{fig:uncertainty_mgp-abc}c. 
The uncertainty is low in honeycomb lattice, as expected, and becomes high around the bi-layer defect, exhibiting also a strong correlation with the coordination number.
Thus, the mapped uncertainty is an accurate automatic indicator of the novelty of the new configurations.
We note that after iterative model refinement by the hierarchical training scheme, the triangular lattice no longer appears in the simulation. 
This example illustrates how the hierarchical active learning scheme helps avoid qualitatively incorrect results that arise due to the a priori unknown relevant configurations. In contrast, a single-pass training of a force field based on manually chosen configurations is very likely to result in non-physical predictions.




To investigate the full phase transition mehanism of stanene, we consider low temperature (below 200K) and high temperature (above 500K) values, corresponding to the crystal and liquid phase for bulk tin. 
In studying the graphite-diamond transition, Khaliullin et al.\cite{khaliullin2011nucleation} identified that nucleation and growth of the disordered phase play key roles in the phase transition process. 
We also find that in stanene the phase transition begins with the appearance of defects and the consequent nucleation and growth of the disordered phase that eventually transforms the structure.

\begin{figure*}[h!]
    \centering
    \includegraphics[]{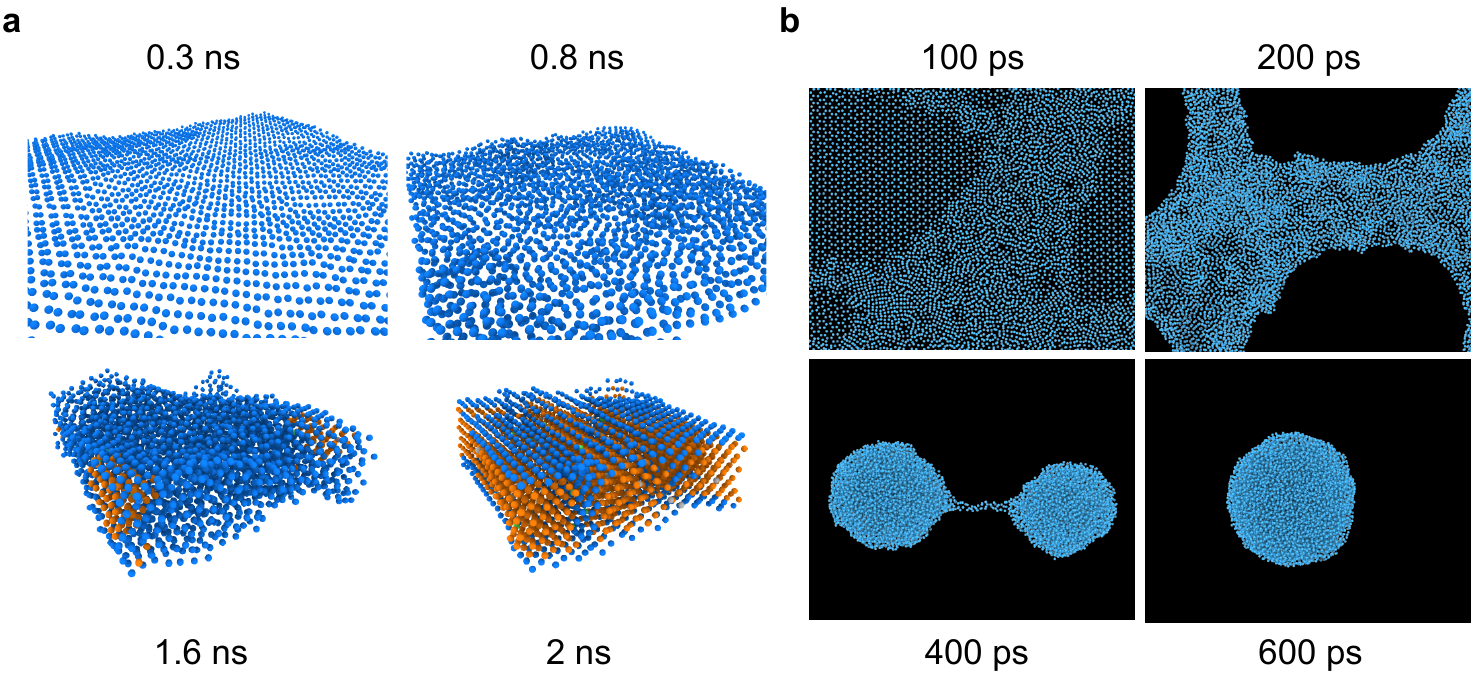}
    \caption{Large-scale molecular dynamics simulation of the inter-dimensional transformation of stanene.
    \textbf{a.} The monolayer - bulk transition at $\sim 200$K. Using common neighbor analysis, the orange atoms are identified as bcc sites, and atoms of other colors are identified as other sites. 
    \textbf{b.} Frames from $\sim 500$K NPT simulation with 10,000 atoms ($39.4 \text{nm} \times 22.7 \text{nm}$) at 100, 200, 400 and 600 ps, showing the melting process of the monolayer.}
    \label{fig:production}
\end{figure*}

From DFT ab-initio calculations, the ground state energy of honeycomb monolayer is $\sim 0.38$ eV/atom higher than the bulk, so the 2D monolayer structure is expected to be metastable. The full transformation described below is therefore a sequence of 2D melting, followed by a kinetically controlled transition to the bulk phase of lower free energy.
We set up the simulation of stanene monolayer of size $15.6 \text{nm} \times 18.1 \text{nm}$ (3000 atoms), and run MD in the NPT ensemble for 3 ns at the temperature of 200K. 
We discover that the ordered honeycomb monolayer loses long-range order and undergoes densification, and finally rearranges into the bcc bulk structure spontaneously, following three transformation steps.
At first, the atoms vibrate around the honeycomb lattice sites, and the monolayer remains in its ordered 2D phase (Fig.\ref{fig:production}a - 0.3 ns). 
The main deviation from the ideal lattice is the out-of-plane rippling due to membrane fluctuations. The fluctuating ripples are similar to what was reported in simulations of graphene \cite{ganz2017initial, PhysRevB.91.045415, zakharchenko2011melting}. 
Next, double-layer stacking defects appear due to buckling and densification. These defect regions grow and quickly expand to the whole system, indicative of a first order phase transition in 2D. The atomic configuration appears to be a ``disordered'' phase (Fig.\ref{fig:production}a - 0.8 ns). In this stage, the structure becomes denser and increasingly three-dimensional driven by energetic gains from increasing Sn atom coordination, with the unit cell dimensions shrinking. 
Finally, patches of bcc lattice nucleate during the atomic rearrangement (Fig.\ref{fig:production}a - 1.6 ns), and those patches grow until the whole system forms the crystal bcc slab and atoms start vibrating around the lattice sites (Fig.\ref{fig:production}a - 2 ns).

The stable phase of bulk tin below 286.35K is expected to be the $\alpha$-phase with the diamond structure, which is different from what we observe here. The reason for the formation of the bcc phase is kinetic rather than thermodynamic, caused by a lower free energy barrier separating the 3D disordered structure from the bcc phase than the $\alpha$-phase, making the former phase more kinetically accessible.
One well-known example is the graphite-diamond transition, in which the meta-stable hexagonal diamond (HD) is obtained from hexagonal graphite (HG) in most laboratory synthesis, instead of the more thermodynamically stable cubic diamond (CD) \cite{bundy1963direct, bundy1996pressure, irifune2003correction, britun2004diffusionless, ohfuji2009origin}. 
Computational study \cite{khaliullin2011nucleation} reveals that the higher in-layer distortions result in higher energy barrier for HG - CD than HG - HD transformation, under most of the experimental conditions.
We observe that the 2D stacking order-disorder and the 3D crystallization transformation steps proceed via nucleation and growth, where the defect nuclei grow and turn the whole system into the intermediate disordered dense phase, while in the second step the bcc nuclei appear from the disordered configuration and grow, turning the system into the bcc crystal phase. 
Our simulation result implies the meta-stability of the free-standing stanene monolayer. The evidences provided by the MD simulation are consistent with the experimental fact that all the existing growths of the stanene monolayer are achieved on substrates \cite{deng2018epitaxial,liao2018superconductivity, baskes1987application, zhu2015epitaxial, pang2020epitaxial, yuhara2018large, gao2016exploring}.
The role that the substrate plays in stabilizing the two-dimensional stanene monolayer is thus of great interests and realistic significance, and will be investigated in our future work.

To investigate the melting behavior at higher temperature, we focus on a 600 ps long MD run that starts from a pristine sample of stanene with size $39.4 \text{nm} \times 22.7 \text{nm}$ (10,000 atoms) at $\sim 500$K. 
A few frames of this simulations are shown in Fig.\ref{fig:production}b. The melting process can also be identified as three stages.
The first stage is the development of large out-of-plane ripples in the free-standing 2D lattice. 
In the second stage, at around 100 ps, double-layer defects in the monolayer form and agglomerate, causing densification. After 200 ps, densification due to large patches of disordered structures lead to the formation of holes, and a large void area grows. 
We observe breaking of the 2-dimensionality of the crystal and formation of stacked configurations where atoms climb over other atoms of 2D stanene. 
In the third stage, the disordered stacked configurations shrink into spherical liquid droplets connected by a neck (400 ps). Subsequently the droplets merge to form a large sphere, minimizing the surface energy (600 ps). In Supplementary Figure 3, we include more discussions and animations of the transition processes.

\subsection*{\label{sec:discussions}Performance}

\begin{figure*}[h!]
    \centering
    \includegraphics{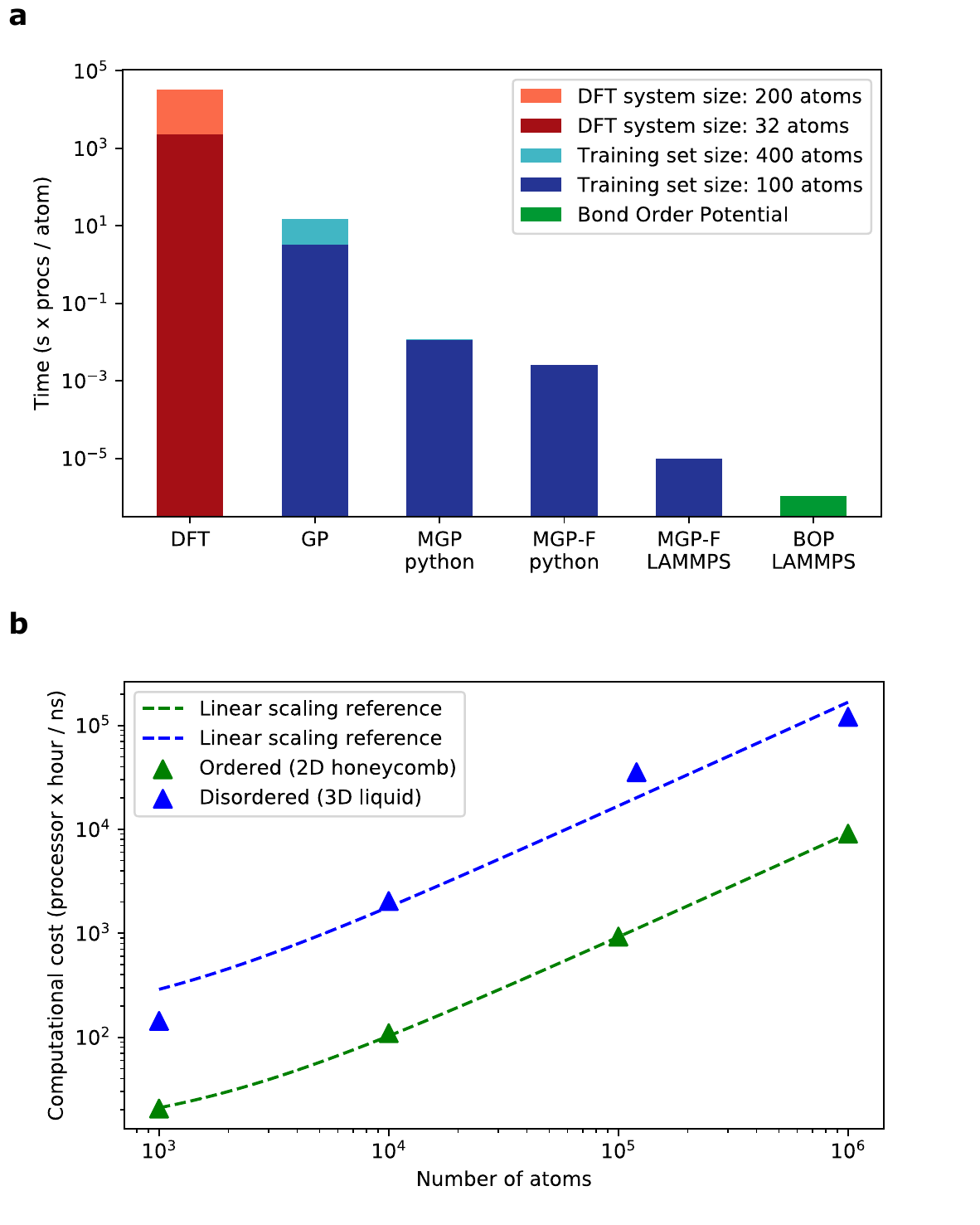}
    \caption{Computational cost and scaling.
    \textbf{a.} The comparison of prediction time ($\text{s}\times \text{processors} / \text{atom}$) between DFT, full GP, MGP (Python, with force and uncertainty), MGP-F (Python, with force only), MGP (LAMMPS, with force only), BOP (LAMMPS). 
    \textbf{b.} The scaling of MGP in LAMMPS with respect to different system sizes, in different phases. Green: 2D honeycomb lattice. Blue: 3D dense liquid.}
    \label{fig:timing}
\end{figure*}

In this section, we discuss the computational cost of the MGP model in application to large scale MD simulations, and its advantages and limitations. Timing results are shown in Fig.\ref{fig:timing}a, comparing different models including DFT, full GP (without mapping; implemented in Python), MGP (with mapped force and uncertainty; Python), MGP-F (MGP with force prediction only; Python), MGP (force only; implemented in LAMMPS) and BOP (LAMMPS force field). 
For DFT, we tested prediction time for system sizes of 32 atoms and 200 atoms, where computational cost grows as the cube of the system size. The other algorithms' cost scales linearly with system size, since they rely on local environments with finite cutoff radius, we measure the same \textit{per atom} computational cost for both system sizes. 

It is important to highlight the acceleration of the MGP relative to the original full GP. 
We present the timings of GP and MGP built from different training set sizes (100 and 400 training data) in Fig.\ref{fig:timing}, where the GP scales linearly. 
MGP-F (without uncertainty) and MGP (with uncertainty, fixed rank) are independent of the training set size. 
In the stanene system MGP BFF is two orders of magnitude faster than GP with $O(10^2)$ training data points when both force and variance are predicted, and the speedup becomes more significant as the GP collects more training data, as reported in Supplementary Figure 4. 
We also note that although the prediction of MGP gets rid of the linear scaling with training set size, the construction of the mappings does scale linearly with training set size, since it requires GP to make prediction on each grid point.

\begin{table}[htbp]
    \centering
    \begin{tabular}{|c|c|}   
        \hline
        System size (atoms) & $1,000,000$\\
        CPU number & 256 \\
        Thermostat & NVT \\
        Temperature (K) & 100 \\
        BOP cutoff $(\text{\AA})$ & $6.0$ \\
        MGP cutoff $(\text{\AA})$ & $7.2$ \\
        BOP speed $(\text{atom}\cdot\text{timestep/processor/s})$ & $\sim 2.3\times 10^6$ \\
        MGP speed $(\text{atom}\cdot\text{timestep/processor/s})$ & $\sim 5.0\times 10^5$ \\
        \hline
    \end{tabular}
    \caption{Performance in LAMMPS}
    \label{tab:timing_1M}
\end{table}

The cost of the MGP force field implemented in LAMMPS (force-only) is comparable to that of empirical interatomic BOP force field, since both BOP and MGP models consider interatomic distances within the local environment of a central atom and thus have similar computational cost at the same cutoff radius. 
For accurate simulations, however, the MGP uses a larger cutoff radius ($7.2 \text{\AA}$) than what is used by the BOP ($6 \text{\AA}$), and thus a simulation with our MGP force field for stanene is slower by a factor of $4 \sim 7$ approximately, see our test of 1 million atoms in Table 1. 
The computational cost of the MGP force field in LAMMPS as a function of the system size is shown in Fig.\ref{fig:timing}b. The prefactor of the linear dependence depends on the number of pairs and triplets within a local environment. The scaling factor for the 2D honeycomb lattice structure is lower than for the dense 3D liquid, because each atom in the 3D liquid has more neighbors than the honeycomb lattice within the same cutoff radius. 

The hierarchical active learning scheme adopted here is an efficient path to systematically improve the force field. However, it must be noted that the training is still limited by the system sizes that can be tackled with a DFT calculation. In fact, DFT calculations with thousands of atoms are unaffordable. 
To circumvent this problem, regions of high uncertainties may be cropped out of the large simulation cell, in order to perform DFT calculation on a smaller subset of atoms with a relevant structure. In this direction, we note that uncertainty is a local predicted property in MGP BFF, indicating how each atom contributes to the total uncertainty. In the Fig.\ref{fig:uncertainty_mgp-abc}c we show that the largest uncertainty in an MD frame of 2048 atoms comes from regions close to defects, since these configurations are missing in the training set for that particular model. Therefore, one may use the uncertainty to automatically select a section of the crystal close to the high-uncertainty region, and use a DFT simulation to add a training data point to the MGP. Automatically constructing such structures remains an open challenge in complex materials. However, access to the local uncertainty for each force prediction already allows for assessment of reliability of simulations, and is central to the paradigm of BFFs. 

Finally, we mention that the descriptive power of our current BFF is limited due to the lack of full many-body interactions in GP kernel, since the $n$-body kernel only uses low-dimensional functions of crystal coordinates. In this work, we have implemented the MGP for 2- and 3-body interactions, which, for the mapping problem, translate to interpolation problems in 3 dimensions. The approach may be extended to higher interaction orders at the expense of computational efficiency, but it becomes expensive in terms of both computation and storage requirements for splines on uniform grids to reach a satisfactory accuracy. Many-body descriptors can be taken into account to increase the descriptive power. To map it in the same way as the methodology we introduced, it is necessary to have a descriptor such that it can be decomposed into low-dimensional functions. For example, the SOAP approach of Refs. \cite{jinnouchi2019phase, jinnouchi2019fly} cannot be mapped using the methodology followed in this work, since it relies on a high-dimensional descriptor. In particular, the interpolation procedure may be too computationally expensive to be a viable solution.


\section*{\label{sec:conclusion}Discussion}
In conclusion, we present an extended method for mapping the Gaussian Process regression model with many-body kernels, where we accurately parameterize both the force and its variance obtained from the original Bayesian model as functions of atomic configurations. 
The mapped forces and their variance are then utilized in an active learning workflow to obtain a machine learned force field that includes its own uncertainty, the first realization of a fast {\it Bayesian force field}. 
The resulting model has near-quantum accuracy and computational cost comparable to empirical analytical interatomic potentials. Large scale simulations are used as a demonstration to investigate the microscopic details of inter-dimensional transformation behavior of 2D stanene into 3D tin. We discover that the transformation proceeds by nucleation of bilayer defects, densification due to continued disordered multilayer stacking, and finally conversion into either crystalline or molten bulk tin, depending on temperature.
This application shows that we can actively monitor the uncertainties of force predictions during molecular dynamics simulations and iteratively improve the model as the material undergoes reorganization, exploring diverse structures in the configuration phase space. The ability to reach simulation sizes of over 1 million atoms is promising for application of machine learning force fields to study phase diagrams and transformation dynamics of complex systems, while automated active learning of force fields opens a way to accelerate wide-range material design and discovery.

\section*{\label{sec:method}Methods}

\subsection*{Background: Gaussian Process Force Field}

In this section, we summarize the GP model introduced in Ref. \cite{vandermause2020fly}.
From here on, we use bold letters to denote matrices/vectors and regular letters for scalars/numbers. 

An \textit{atomic configuration} is defined by atomic coordinates $\boldsymbol{R}_\alpha$, a vector of dimension equal to the number of atoms $N_{atoms}$, where $\alpha$ labels the three Cartesian directions $\alpha=x, y, z$.
While in general the crystal may contain atoms of different species, we here consider only a single species case for the sake of simplicity. The results can be extended to the multi-component case, which is discussed in Supplementary Method 1.

The objective of the GP regression model is to predict the forces $\boldsymbol{F}_\alpha$ for an given atomic configuration $\boldsymbol{R}_\alpha$.
To this aim, we associate the forces of an atom $i$, $F_{i, \alpha}$, to the local environment around this atom $\env_i$.
$F_{i, \alpha}$ can then be computed by comparing its environment with a training data set of atomic environments with known forces, which is quantified by distance $d$ between the target environment and the training set.

The \textit{atomic environment} $\env_i$ of the atom $i$ is defined as all the atoms that are within a sphere of a cutoff radius $r_{cut}$ centered at the atom $i$. 
We proceed by partitioning this atomic environment $\env_i$ into a set of $n$-atom subsets, denoted by $\env^{(n)}_i$. 
A distance $d^{(n)}$ can be defined to quantify the difference between environments. 
We show examples of $n=2$ and $n=3$ on the construction of $\env^{(n)}_i$ and $d^{(n)}$.
For $n=2$, $\env^{(2)}_i$ is a set of pairs $p$, between central atom $i$ and another atom $i'$ in $\env_i$,
\begin{equation}
    \env^{(2)}_i= \{p=(i,i'); i' \in \env_i, i' \neq i\};
\end{equation} 
To each atomic pair $p$ we associate an interatomic distance $r_p$ between the two atoms.
The distance metric between two pairs $p$ and $q$ is then defined as $(r_p-r_q)^2$.
And the distance metric between two atomic environments $\env_i^{(2)}$ and $\env_j^{(2)}$ is defined as
\begin{equation}
(d^{(2)}_{ij})^2 = \sum_{p \in \env_i^{(2)}, q \in \env_j^{(2)}}(r_p - r_q)^2.
\end{equation}
This distance $d^{(2)}$ will be used to determine the two-body contribution to the atomic forces, for the purposes of constructing the kernel function.

This construction can be extended to higher order interatomic interactions.
In this work, we also include three-body interactions, corresponding to $n=3$.
Similar to the two-body case, the set $\env_i^{(3)}$ of triplets $p$ between the central atom $i$ and two other atoms $i'$ and $i''$ in $\env_i$ is defined as, 
\begin{equation}
    \env^{(3)}_i= \{p=(i,i',i''); i', i'' \in \env_i, i' \neq i, i'' \neq i, i'\neq i'' \}\;.
\end{equation}
This time, each triplet $p$ is characterized by a vector $\boldsymbol{r}_p=\{r_{p,1}, r_{p, 2}, r_{p,3}\}$, which represents the three interatomic distances of this atomic triplet $p$.
Including permutation, the distance metric between two triplets $p$ and $q$ can be defined as
$\sum_{u, v} (r_{p, u} - r_{q,v})^2$.
And the 3-body distance metric between two atomic environments $\env_i^{(3)}$ and $\env_j^{(3)}$ is
\begin{equation}
    (d^{(3)}_{ij})^2 =  \sum_{p \in \env_{i}^{(3)}, q \in \env_{j}^{(3)}} \sum_{u, v} (r_{p, u} - r_{q, v})^2\;.
\end{equation}

Having now defined a distance between atomic environments, we can build a GP regression model for atomic forces.
We define the equivariant force-force kernel function to compare two atomic environments as
\begin{align}
k_{\alpha \beta}(\env_i, \env_j) 
=
\sum_n
k^{(n)}_{\alpha \beta}(\env^{(n)}_i, \env^{(n)}_j) 
= 
\sum_n
\frac{\partial^2 k^{(n)}(\env^{(n)}_i, \env^{(n)}_j)}{\partial R_{i\alpha} \partial R_{i\beta}} \;,
\label{eq:derv}
\end{align}
where $\alpha$ and $\beta$ label Cartesian coordinates, and $\frac{\partial}{\partial R_{i\alpha}}$ indicates the partial derivative of the kernel with respect to the position of atom $i$.
The $n$-body energy kernel function $k^{(n)}(\env^{(n)}_i, \env^{(n)}_j)$ is first expressed in terms of the partitions of $n$ atoms of the two atomic environments $\env_i$ and $\env_j$, and is defined as:
\begin{align}
\label{eq:n-body-kernel}
k^{(n)}(\env_i^{(n)}, \env_j^{(n)})
=
\sum_{\substack{p \in \env_i^{(n)}\\ q \in \env_j^{(n)}}} \tilde{k}^{(n)}(p, q) 
= 
 \sum_{\substack{p \in \env_i^{(n)}\\ q \in \env_j^{(n)}}} \sigma_\mathrm{sig}^{(n)}
\exp \bigg[ - \frac{ (d^{(n)}_{ij})^2 }{ 2 (l^{(n)})^2 } \bigg]\varphi_\mathrm{cut}^{(n)}
\;,
\end{align}
where $\tilde{k}^{(n)}(p, q)$ is a Gaussian kernel function, built using the definition of distance between $n$-atoms subsets as described before.
The $\sigma_\mathrm{sig}^{(n)}$ and $l^{(n)}$ are hyperparameters independent of $p$ and $q$,  that set the signal variance related to the maximum uncertainty and the length scale, respectively.
The smooth cutoff function $\varphi_\mathrm{cut}^{(n)}$ dampens the kernel to zero as interatomic distances approach $r_\mathrm{cut}$. 
In this work, we chose a quadratic form for the cutoff  function,
\begin{align}
\varphi_\mathrm{cut}^{(2)}(p, q) &= (r_q-r_\mathrm{cut})^2(r_p-r_\mathrm{cut})^2, \\
\varphi_\mathrm{cut}^{(3)}(p, q) &= \prod_{s} (r_p^s-r_\mathrm{cut})^2 \prod_{t} (r_q^t-r_\mathrm{cut})^2.
\end{align}

The GP regression model uses training data as a reference set of $\Ntrain$ atomic environments.
In particular, predictions of force values and variances require a $3\Ntrain\times 3\Ntrain$ covariance matrix 
\begin{equation}
\boldsymbol{\covariance} 
= \covariance_{i\alpha,j\beta}
= k_{\alpha \beta}(\env_i,\env_j)+\sigma_n^2\delta_{ij}\delta_{\alpha \beta}
\label{eq:cov_matr}
\end{equation}
where $\sigma_n$ is a noise parameter (indices $i \alpha$ and $j \beta$ are grouped together). 
We define an auxiliary $3\Ntrain$ dimensional vector $\boldsymbol{\eta}$ with components 
\begin{equation}
\eta_{i\alpha} := (\covariance^{-1})_{i\alpha, j\beta} F_{j\beta}\;.
\label{eq:eta}
\end{equation} 

In order to make a prediction of forces and their errors on a target structure $\env_i$, we compute the Gaussian kernel vector $\bar{\boldsymbol{k}}_{i\alpha}$ for atom $i$ and a direction $\alpha$ ($\alpha=x,y,z$), where the components of the vector are kernel distances $(\bar{k}_{i\alpha})_{j\beta } := k_{\alpha\beta}(\env_i,\env_j)$ between the prediction $i\alpha$ and the training set points $j\beta$, $j=1,...,\Ntrain$, $\beta=x,y,z$. 
Finally, the predictions of the mean value of the Cartesian component $\alpha$ of the force $F_{i\alpha}$ acting on the atom at the center of the atomic environment $\env_i$ and its variance $V_{i\alpha}$ are given by \cite{rasmussen2003Gaussian}:
\begin{align}
F_{i\alpha}&= \bar{\boldsymbol{k}}_{i\alpha}^\top\boldsymbol{\eta}
\label{eq:force}  \;, \\
V_{i\alpha} &= k_{\alpha \alpha}(\env_i,\env_i) - 
\bar{\boldsymbol{k}}_{i\alpha}^\top 
\boldsymbol{\covariance}^{-1}
\bar{\boldsymbol{k}}_{i\alpha}
 \;. 
\label{eq:var}
\end{align}

We notice that this formulation has two major computational bottlenecks, that we intend to address in this work.
First, the vector $\bar{\boldsymbol{k}}_{i\alpha}$ needs to be constructed for every force prediction; for each atom, the evaluation of $\bar{\boldsymbol{k}}_{i\alpha}$ requires the evaluation of the kernel between the atomic environment and the entire training data set, i.e. $\Ntrain$ calls to the Gaussian kernel. We discuss an approach to speed up the evaluation of $\bar{\boldsymbol{k}}_{i\alpha}$. 
Second, every variance calculation requires a matrix-vector multiplication that is computationally expensive, and thus will be further discussed in the \textit{Methods} section.


\subsection*{Mapped Force Field\label{subsec:map-force}}

In GP regression, a kernel function is used to build weights that depend on the distance from training data: training data that are similar to the test data contribute with a larger weight, and vice versa, dissimilar training data contributes less.
To make a prediction on a data point (force on an atom in its local environment), the GP regression model requires the evaluation of the kernel between such point and all the $\Ntrain$  training set data points.
The GP mean prediction is a weighted average of the training labels, using the distances as measured by the kernel function as \textit{weights}.
In Eq. \ref{eq:force}, we notice that the predictive mean requires the evaluation of a $3\Ntrain$-dimensional vector  $\bar{\boldsymbol{k}}_{i\alpha}$, which quantitatively weighs the contribution to the test atomic environment $\env_i$ coming from the training atomic environments $\env_j$.

In the following discussion, we consider a specific subset size $n$, capturing $n$-body atomic structure information, i.e. $(\bar{k}^{(n)}_{i\alpha})_{j\beta}=k^{(n)}_{\alpha \beta}(\env^{(n)}_i, \env^{(n)}_j) $ without loss of generality. The same treatment below can be implemented on each $n$ and summed over the mappings of all ``$n$"s to get the final prediction if we are using a kernel function of the form of Eq.\ref{eq:derv}.

Starting with the idea introduced in Ref. \cite{glielmo2018efficient}, the kernel definition (Eq.\ref{eq:n-body-kernel}) allows us to decompose the force prediction into contributions from $n$-atom subsets, and to decompose the variance into contributions from pairs of $n$-atom subsets.

To see this, we first note that the kernel vector of atomic environments $\env_i$ can be decomposed into the kernel of all the $n$-atom subsets in $\env_i$, and the kernel of an $n$-atom subset $p$ can be further decomposed into the contributions from all the $n-1$ neighbor atoms (except for the center atom $i$). 
We denote $\mathbf{r}_p^s:=\mathbf{R}^s-\mathbf{R}^i$ as a bond in $p$, where $s$ labels the remaining $n-1$ neighbor atoms, and introduce the unit vector $\hat{r}_p^s:=\mathbf{r}_p^s/r_p^s$ denoting the direction of the bond. Then the force kernel has decomposition
\begin{align}
\bar{\boldsymbol{k}}^{(n)}_{i\alpha} 
= \sum_{p\in \env_i} \tilde{\boldsymbol{k}}^{(n)}_{i\alpha}(p) 
= \sum_{p} \sum_{s} \boldsymbol{\mu}^{(n)}_{s}(p) \hat{r}_{p,\alpha}^s \;,
\label{eq:force_kern_decomp}
\end{align}
where
\begin{align}
\left[\tilde{k}^{(n)}_{i\alpha}(p)\right]_{j\beta}
:= \frac{\partial^2 \tilde{k}^{(n)}(p,q)} {\partial R^i_{\alpha} \partial R^j_{\beta} }
\end{align}
and
\begin{align}
\Big[\mu^{(n)}_{s} (p) \Big]_{j\beta}    
:=  \sum_{q \in \env_j} \sum_{t}     \frac{\partial^2 \tilde{k}^{(n)}(p, q)}{\partial r_p^s \partial r_q^t} \hat{r}_{q,\beta}^{t}
\end{align}
are vectors in $\mathbb{R}^{3\Ntrain}$ space. 
In the first equality of Eq.\ref{eq:force_kern_decomp} we used the decomposition of Eq. \ref{eq:n-body-kernel}, and in the second equality we first convert the derivative from atomic coordinates to the relative distances of the $n$-atom subset, then take advantage of the fact that the kernel only depends on relative distances between atoms, i.e. $r_p^s=\|\mathbf{r}_p^s\|$.

Therefore, forces may be written as
\begin{align}
F^{(n)}_{i\alpha}
=\sum_{p \in \env_i}  \sum_{s} 
    f^{(n)}_s(p) 
    \hat{r}^s_{p,\alpha}
    \;.
    \label{eq:force-decomp}
\end{align}
where
\begin{equation}
f^{(n)}_s(p):={\boldsymbol{\mu}^{(n)}_s}^\top (p) \boldsymbol{\eta}
\label{eq:wf-force}
\end{equation} 
and $\boldsymbol{\eta}$ is define in Eq.\ref{eq:eta}. 

As noted in Ref. \cite{glielmo2018efficient}, we can achieve a better scaling by a computationally efficient parametric method, i.e. an approximate way for constructing such functions $f$.
In fact, a parametric function can quickly yield the value without evaluating kernel distances from all the GP training points,
making the cost independent of the size of the training set, thus remove a computational bottleneck of the full GP implementation.

In order to achieve an effective parametrization of these weight functions, we use a cubic spline interpolation on a \textit{uniform grid} of points, which benefits from having an analytic form and a continuous second derivative.
By increasing the number of points in the uniform grid, the MGP BFF prediction can be made arbitrarily close to the original functions.

The construction of the mapped force field includes the following steps:

\textbf{Step 1}: Define a uniform grid of points $\{x_k\in [a,b]^{\Nsubset}\}$ in the space of distances given by the vectors $\mathbf{r}_p$. 
The lower bound $a\geq 0$ is chosen to be smaller than the minimal interatomic distance, and the upper bound $b$ can be set to the cutoff radius of atomic neighborhoods, above which the prediction vanishes (using the zero prior of GP), 
and $\Nsubset=\frac{n(n-1)}{2}$ is the number of interatomic distances in the $n$-atom subset.

\textbf{Step 2}: A GP model with a fixed training set yields the values $\{f^{(n)}_s(x_k)\}$ for each grid point $x_k$ using Eq.\ref{eq:wf-force}. 

\textbf{Step 3}: Interpolate with a cubic spline function $\hat{f}^{(n)}_s$ through the points $\{(x^{(n)}_k, f^{(n)}_s(x_k))\}$.\\

In prediction, contributions from all $n$-bodies are added up, so the force of local environment $\rho$ is predicted as:
\begin{equation}
    \hat{F}_\alpha(\env) = \sum_n \sum_{p \in \env^{(n)}} \sum_{s} \hat{f}^{(n)}_s(p) \hat{r}_{p,\alpha}^s  \nonumber \;.
\end{equation}

\subsection*{Mapped Variance Field\label{subsec:map-var}}

Using a derivation similar to that of the force decomposition (Eq.\ref{eq:force-decomp}), we derive the decomposition of the variance
\begin{align}
V_{i\alpha} 
&= \sum_{n, n'} \sum_{\substack{p \in \env^{(n)}\\ q \in \env^{(n')}}} \sum_{\substack{s, t}} v^{(n, n')}_{s,t}(p,q) \hat{r}^s_{p,\alpha} \hat{r}^t_{q,\alpha} \;,
\label{eq:decomp-v}
\end{align}
where
\begin{align}
v^{(n, n')}_{s,t}(p,q) 
:= 
\delta_{nn'}\tilde{k}^{(n)}_{\alpha \alpha}(p,q)
-
\boldsymbol{\mu}^{(n)}_s(p)^\top \boldsymbol{\covariance}^{-1} \boldsymbol{\mu}^{(n')}_t(q)
\;. \label{eq:wf-var}
\end{align}
with the covariance matrix $\covariance$ defined in Eq.\ref{eq:cov_matr}.

We note that the domain of the variance function $v$ has dimensionality twice that of $f$, so that when the interaction order $n>3$, the number of grid points becomes  prohibitively large to efficiently map $v$.
Even for a 3-body kernel, when $v$ is a 6-D function, we find that the evaluation of the variance, both in terms of computational time and memory footprint, limits its usage to simple benchmarks.

An efficient evaluation of the variance is critical for adopting Bayesian active learning.
We now discuss a key result of this work and introduce an accurate yet efficient approximate mapping field for the variance.
In particular, we focus on simplifying the vector $\boldsymbol{\mu}^{(n)}_s(p)$, which, from our tests, is the most computationally expensive term for predicting variance.
In particular, since $\boldsymbol{\mu}^{(n)}_s(p)$ evaluates the kernel function between the test point $p$ and all the training points, it scales linearly with the training set size.
By implementing a mapping of the variance weight field, the cost of variance prediction depends only on the rank of our dimension reduction approach, and is independent of the training data set size.

To this aim, we first define the vector $\boldsymbol{\psi}_s^{(n)}(p):=\mathbf{L}^{-1} \boldsymbol{\mu}_s^{(n)}(p)\in \mathbb{R}^{3\Ntrain}$, where $\mathbf{L}$ is the Cholesky decomposition of  $\boldsymbol{\covariance}$, i.e. $\boldsymbol{\covariance}=\mathbf{L} \mathbf{L}^\top$, so that Eq. \ref{eq:wf-var} can be written as 
\begin{equation}
v^{(n, n')}_{s,t}(p,q) = \delta_{nn'}\tilde{k}^{(n)}_{\alpha,\alpha}(p,q) - \boldsymbol{\psi}^{(n)}_s(p)^\top \boldsymbol{\psi}^{(n')}_t(q) \;.
\label{eq:small-v}
\end{equation}
The domain of vector function $\boldsymbol{\psi}^{(n)}_s(p)$ is half the dimensionality of that of $v_{s,t}^{(n, n')}(p,q)$. 

As a first idea, since each component of $\boldsymbol{\psi}^{(n)}$ has domain of lower dimensionality, one may think of building $3\Ntrain$ spline functions to interpolate the $3\Ntrain$ components separately, which means the number of spline functions needed grows with the training set size.

To achieve a more efficient mapping of $\boldsymbol{\psi}_s^{(n)}$, we take advantage of the principal component analysis (PCA), a common dimensionality reduction method. The construction of the mapped variance field includes the following steps:

\textbf{Step 1}: We start by evaluating the function values on the uniform grid $\{x_k\in [a,b]^{\Nsubset}, k=1,\dots, G\}$ as introduced in the section above. 
At each grid point $x_k$, we evaluate the vectors $\boldsymbol{\psi}^{(n)}_s(x_k)$ and use them to build the matrix $\mathbf{M}^{(n)}_s := (\boldsymbol{\psi}^{(n)}_s(x_1), \boldsymbol{\psi}^{(n)}_s(x_2), ..., \boldsymbol{\psi}^{(n)}_s(x_G))^\top \in \mathbb{R}^{G\times 3\Ntrain}$. 

\textbf{Step 2}: In PCA we perform singular value decomposition (SVD) on $\mathbf{M}^{(n)}_s$, such that $\mathbf{M}^{(n)}_s = \mathbf{U}^{(n)}_s \mathbf{\Lambda}^{(n)}_s (\mathbf{V}^{(n)}_s)^\top$, where $\mathbf{U}^{(n)}_s\in \mathbb{R}^{G\times 3\Ntrain}$ and  $\mathbf{V}^{(n)}_s\in \mathbb{R}^{3\Ntrain\times 3\Ntrain}$ have orthogonal columns, and $\mathbf{\Lambda}^{(n)}_s\in \mathbb{R}^{3\Ntrain\times 3\Ntrain}$ is diagonal.

\textbf{Step 3}: We reduce the dimensionality by picking the $m$ largest eigenvalues of $\mathbf{\Lambda}^{(n)}_s$ and their corresponding eigenvectors from $\mathbf{U}^{(n)}_s$ and $\mathbf{V}^{(n)}_s$ (here we assume that eigenvalues are ordered).
Then, we readily obtain a low-rank approximation as 
\begin{equation}
\mathbf{M}^{(n)}_s\approx \bar{\mathbf{M}}^{(n)}_s=\bar{\mathbf{U}}^{(n)}_s\bar{\mathbf{\Lambda}}^{(n)}_s(\bar{\mathbf{V}}^{(n)}_s)^\top \;, 
\end{equation} 
with $\bar{\mathbf{\Lambda}}^{(n)}_s\in \mathbb{R}^{m\times m}$, $\bar{\mathbf{U}}^{(n)}_s\in \mathbb{R}^{G\times m}$, and $\bar{\mathbf{V}}^{(n)}_s\in \mathbb{R}^{3\Ntrain\times m}$. 

\textbf{Step 4}: Define $\boldsymbol{\phi}^{(n)}_s:=(\bar{\mathbf{V}}^{(n)}_s)^\top\boldsymbol{\psi}^{(n)}_s$, which is a vector in $\mathbb{R}^{m}$, and interpolate this vector between its values at the grid points  $\{x_k\}$ (in detail, $\phi^{(n)}_{s,i}(x_k) = \bar{U}_{s,ij}^{(n)} \bar{\Lambda}_{s,j}^{(n)}$, for $i,j=1, ..., m$).
As done previously for the force mapping, spline interpolation $\hat{\boldsymbol{\phi}}$ can now be used for approximating the $\boldsymbol{\phi}^{(n)}_s$ vector as well. \\

Therefore, the MGP prediction of the variance is given by
\begin{align}
    \hat{v}^{(n, n')}_{s,t}(p,q)
    :=
    \delta_{nn'}\tilde{k}^{(n)}_{\alpha,\alpha}(p,q) 
    - 
    \hat{\boldsymbol{\phi}}_s^{(n)}(p)^\top (\bar{\mathbf{V}}_s^{(n)})^\top \bar{\mathbf{V}}_t^{(n')}\hat{\boldsymbol{\phi}}_t^{(n')}(q) \;.
    \label{eq:small-v-mgp}
\end{align}
In this form, the variance would still require a multiplication between two matrices of size $m \times 3\Ntrain$.
However, we note that the computational cost of total variance can be reduced
\begin{align}
    \hat{V}_\alpha(\env) 
    =
    k_{\alpha,\alpha}(\env, \env)  
    -\bigg( \sum_{n} \sum_{p \in \env^{(n)}} \sum_{s} \bar{\mathbf{V}}_s^{(n)}\hat{\boldsymbol{\phi}}_s^{(n)}(p) \hat{r}^s_{p,\alpha} \bigg)^2
    \;.
    \label{eq:tot-var}
\end{align}
Nominally, the calculation of the second term in the variance still scales with the data set size $\Ntrain$, since the size of $\bar{\mathbf{V}}^{(n)}$ depends on $\Ntrain$.
However, we find the cost of the vector-matrix multiplication negligible compared to evaluations of the self-kernel function (the 1st term in Eq.\ref{eq:small-v-mgp}).
Therefore, we have formulated a way to significantly speed up the calculation of the variance, which only weakly depends on the training set size.
In addition, by tuning the PCA rank $m$, we can optimize the balance between accuracy and efficiency,  increasing the flexibility of this model. 

The convergence of the accuracy of mapped forces and uncertainty with different grid numbers $G$ and ranks $m$ is discussed in Supplementary Figure 2.


\section*{Data Availability}
The code used to generate results in this paper, and the related data are published in Materials Cloud Archive \cite{talirz2020materials} with 
\href{https://doi.org/10.24435/materialscloud:qg-99}{DOI: 10.24435/materialscloud:qg-99} \cite{data}.

\section*{Code Availability}
The algorithm for the MGP and mapped Bayesian active learning has been implemented in the open-source package FLARE \cite{vandermause2020fly} and is publicly available online: \url{https://github.com/mir-group/flare}. 

The python scripts for generating the stanene force field and MD simulations are also publicly available online: \url{https://github.com/YuuuXie/Stanene_FLARE}.

\section*{Acknowledgement}
We thank Steven Torrisi and Simon Batzner for help with the development of the FLARE code. We thank Anders Johansson for the development of the KOKKOS GPU acceleration of the MGP LAMMPS pair style. We thank Dr. Nicola Molinari, Dr. Aldo Glielmo, Dr. Claudio Zeni, Dr. Sebastian Matera, Dr. Christoph Scheurer, Dr. Nakib Protik and Jenny Coulter for useful discussions. Y.X. is supported by the US Department of Energy (DOE) Office of Basic Energy Sciences under Award No. DE-SC0020128. L.S. is supported by the Integrated Mesoscale Architectures for Sustainable Catalysis (IMASC), an Energy Frontier Research Center funded by the US Department of Energy (DOE) Office of Basic Energy Sciences under Award No. DE-SC0012573. A.C. is supported by the Harvard Quantum Initiative. J.V. is supported by Robert Bosch LLC and the National Science Foundation (NSF), Office of Advanced Cyberinfrastructure, Award No. 2003725.

\section*{Author contributions}
Y.X. developed the mapped variance method. J.V. led the development of the FLARE Bayesian active learning framework, with contributions from L.S. and Y.X. Y.X. implemented the Bayesian active learning training workflow based on the mapped Bayesian force field. L.S. and Y.X. developed the LAMMPS pair style for the mapped force fields. Y.X. trained the models for stanene and performed molecular dynamics and analysis. A.C. guided the setup of DFT calculations. B.K. initiated and supervised the work and contributed to algorithm development. Y.X. wrote the manuscript, with contributions from all authors.

\section*{Competing interests}
The authors declare no competing financial or non-financial interests.

\bibliographystyle{naturemag}
\bibliography{main}

\end{document}


\title{Bayesian Force Fields from Active Learning for Simulation of Inter-Dimensional Transformation of Stanene}
\author{Yu Xie}
\email{xiey@g.harvard.edu}
\affiliation{John A. Paulson School of Engineering and Applied Sciences, Harvard University, Cambridge, MA 02138, USA}

\author{Jonathan Vandermause}
\affiliation{John A. Paulson School of Engineering and Applied Sciences, Harvard University, Cambridge, MA 02138, USA}

\author{Lixin Sun}
\affiliation{John A. Paulson School of Engineering and Applied Sciences, Harvard University, Cambridge, MA 02138, USA}

\author{Andrea Cepellotti}
\affiliation{John A. Paulson School of Engineering and Applied Sciences, Harvard University, Cambridge, MA 02138, USA}

\author{Boris Kozinsky}
\email{bkoz@seas.harvard.edu}
\affiliation{John A. Paulson School of Engineering and Applied Sciences, Harvard University, Cambridge, MA 02138, USA}
\affiliation{%
Robert Bosch LLC, Research and Technology Center,
Cambridge, Massachusetts 02142, USA}


\maketitle

\begin{center}
    \LARGE Supplementary Materials
\end{center}

\toccontents

\newpage
\section*{Supplementary Methods}
\subsection*{Supplementary Method 1: Multi-Component Systems\label{supp-sec:multi-comp}}

In the main text, we derived the mathematical formulation in the context of single component.  We note that our GP and MGP are natural to extend to multiple components, by differentiating combinations of species. 
As described in the main text, the $n$-body kernel compares two $n$-atom subsets $p\in \env^{(n)}_i$, $q\in \env^{(n)}_j$. 
When there are multiple species in the system, we define the kernel function $\tilde{k}^{(n)}(p,q)$ to be non-zero only when $p$ and $q$ have the same species in their corresponding atoms. 
To formulate rigorously, denote $\mathbf{s}_p$ is the sequence of species of atom in $p$, we define the multi-component kernel as

\begin{equation}
    \tilde{k}^{(n)}(p,q)=\delta_{\mathbf{s}_p,\mathbf{s}_q} \exp \bigg( - \frac{ d^{(n)}(p,q)^2 }{ 2 (l^{(n)})^2 } \bigg)\varphi_\mathrm{cut}^{(n)}(p, q)\;,
\end{equation}

Therefore, the mapped force and variance can be extended to multi-component system in the way that for each possible $\mathbf{s}_p$, we construct $\hat{f}^{(n)}(p)$ and $\hat{\boldsymbol{\phi}}^{(n)}(p)$ the same as the single-component system that is described in the main text. In prediction, the force and variance contributions from $n$-atom subsets in the testing atomic environment are predicted by the spline functions corresponding to their species.

\newpage
\section*{Supplementary Tables}
\subsection*{Supplementary Table \ref{supp-tab:gp_hyps}: On-the-fly Training Settings of Gaussian Process}

\begin{table}[htbp]
    \centering
    \begin{tabular}{|c|c|ccc|ccc|}
        \hline
        \multirow{2}{*}{Kernel} & \multirow{2}{*}{Cutoff ($\text{\AA}$)}
        & \multicolumn{3}{c|}{Initial Hyps}
        & \multicolumn{3}{c|}{Optimized Hyps}\\
        \cline{3-8}
        & & $\sigma_\mathrm{sig}$ (\eVperA) & $l$ ($\text{\AA}$) & $\sigma_n$ (\eVperA) & $\sigma_\mathrm{sig}$ (\eVperA) & $l$ ($\text{\AA}$) & $\sigma_n$ (\eVperA) \\
        \hline 
        2-body & 7.2 & 0.2 & 1.0 & 0.05 & 3.27e-2 & 0.518 & 0.140 \\
        \hline
        3-body & 7.2 & 1.0e-4 & 1.0 & 0.05 & 3.75e-5 & 0.691 & 0.140 \\
        \hline
    \end{tabular}
    \caption{GP hyperparameters}
    \label{supp-tab:gp_hyps}
\end{table}

In the Bayesian active learning of stanene, we start constructing a $4\times 4\times 1$ supercell of stanene (32 atoms in total), a size that can be tackled with first-principles techniques. 
The lattice constant $a=4.58 \text{\AA}$ and buckling height $d=0.83 \text{\AA}$ are determined from the DFT structural optimization. The vacuum size of $c=20 \text{\AA}$ is used to make sure that the stanene monolayer is well isolated so that the periodic boundary condition in the z-direction has little influence on the result. 
We perform MD with the MGP BFF using a time step of 1 fs, and the Velocity Verlet algorithm to integrate Newton's equations of motion, in the NVE ensemble.

We use Quantum-ESPRESSO \cite{giannozzi2009quantum}, a plane-wave implementation of DFT, and an ultra-soft pseudo-potential with PBEsol \cite{csonka2009assessing} exchange-correlation functional for Sn. The k-point mesh is $4\times 4 \times 1$, and the energy cutoff is $50$ Ry, with charge density cutoff of $400$ Ry. 

Python is used in all our code except for DFT. FLARE \cite{vandermause2020fly} is our base framework of Bayesian active learning with GP, and the newly developed MGP is inserted into the workflow. 
The radial cutoffs for local environments are chosen as $7.2 \text{\AA}$, the right end of the fourth peak in the radial distribution function (RDF). 
The cutoffs are tuned from maximizing the objective function (maximize likelihood and minimize redundancy). 

The initialization of signal strength ($\sigma$), length scale $l$ and noise $\sigma_n$ are shown in the table below, and are optimized to maximize the marginal likelihood function during the training, using the BFGS algorithm \cite{broyden1970convergence, fletcher1970new, goldfarb1970family, shanno1970conditioning}. More details on the hyperparameter optimization are explained in our earlier work \cite{vandermause2020fly}.

\newpage
\subsection*{Supplementary Table \ref{supp-tab:mgp_hyps}: On-the-fly Training Settings of Mapped Gaussian Process}

\begin{table}[htbp]
    \centering
    \begin{tabular}{|c|c|c|c|}
        \hline
         & grid number & lower bound & upper bound\\
        \hline
        2-body & 144 & $2.35\ \text{\AA}$ & $7.2\ \text{\AA}$ \\
        \hline
        3-body & $128\times 128\times 128$ & $2.35\ \text{\AA}$, $2.35\ \text{\AA}$, $0\ \text{rad}$ & $7.2\ \text{\AA}$, $7.2\ \text{\AA}$, $\pi\ \text{rad}$ \\
        \hline
    \end{tabular}
    \caption{MGP hyperparameters}
    \label{supp-tab:mgp_hyps}
\end{table}

For MGP, the hyperparameters are shown in Table \ref{supp-tab:mgp_hyps}. To interpolate a function, we need to choose a closed interval and determine the grid points in this interval. 
The number of grid points for the mapped spline models is chosen based on the trade-off between accuracy and efficiency of the interpolation. 
For 2-body, the interval is chosen as $[a,b]$, where $a$ should be smaller than the minimal interatomic distance in MD simulation, and $b$ is set to be the same as the 2-body cutoff of GP. 
For 3-body, we choose a ``cube'' space, where we find a closed interval in each dimension. Since 3-body describes triplets, i.e. the triangle formed by three atoms, that can be determined by the two bonds connected to the center atom, $r_1$, $r_2$, and the angle $a_{12}=\langle\hat{r}_1, \hat{r}_2\rangle$ between $r_1$ and $r_2$, we use $(r_1, r_2, a_{12})$ as grid points to build up interpolation function. So the interval for $r_1$ and $r_2$ should be chosen the same as 2-body, only the upper bound is changed to the 3-body cutoff of GP. For the angle, the lower bound and upper bound are fixed to be $0$ and $\pi$, which are not varied in different systems and not chosen by users.

The 1-D and 3-D cubic spline functions come from Github repository \textit{interpolation.py}  \cite{econforge}.

\newpage
\subsection*{Supplementary Table \ref{supp-tab:bulk_properties}: Bulk Polymorphism}

\begin{table}[htbp]
    \centering
    \begin{tabular}{
    |>{\centering\arraybackslash}p{1.5cm}
    |>{\centering\arraybackslash}p{5cm}
    |>{\centering\arraybackslash}p{2cm}
    >{\centering\arraybackslash}p{2cm}
    >{\centering\arraybackslash}p{2cm}
    >{\centering\arraybackslash}p{2.5cm}|
    }
        \hline
        Phase & Property & $\text{MGP}_{f-label}$ & $\text{MGP}_{e-label}$ & DFT & Experiment \\
        \hline
        & Lattice constant $a$ ($\text{\AA}$) & 6.44 & 6.48 & 6.48 & $6.491^a$, $6.483^b$ \\
        $\alpha$-Sn & Bulk modulus (GPa) & 79.46 & 40.90 & 41.86 & $42.6^c$ \\
        & Force MAE (\eVperA) & 0.024 & 0.043 & - & -\\
        \hline
        & Lattice constant $a$ ($\text{\AA}$) & 6.06 & 5.80 & 5.82 & $5.820^a$, $5.831^b$ \\
        \multirow{2}{*}{$\beta$-Sn} & Lattice constant $c/a$ & 0.541 & 0.552 & 0.546 & $0.546^b$\\
        & Bulk modulus (GPa) & 51.42 & 57.36 & 56.97 & $57.0^c$, $54.9^d$ \\
        & Force MAE (\eVperA) & 0.034 & 0.053 & - & -\\
        \hline 
    \end{tabular}
    \caption{a. Ref.\cite{wyckoff1963interscience}, b. Ref.\cite{barrett1980structure}, c. Ref.\cite{brandes2013smithells}, d. Ref.\cite{vaboya1970compressibility}}
    \label{supp-tab:bulk_properties}
\end{table}

We also examine the ability of active learning and BFFs to capture the bulk polymorphism of Sn by collecting training data from $\alpha$-Sn and $\beta$-Sn, and then computing the physical properties such as lattice constant, bulk modulus and phonon dispersion with the resulting force field.
We compare the performance of two models with different training data. The first model $\text{MGP}_{f-label}$ uses the training data of 872 force labels, and the second model $\text{MGP}_{e-label}$ also includes 10 frames of global energies for training.
The lattice constant, bulk modulus and mean absolute error of force prediction obtained from different methods are compared in Table \ref{supp-tab:bulk_properties}. We also computed the phonon dispersion of $\alpha$-Sn and $\beta$-Sn, and the results are shown in Supplementary Figure \ref{supp-fig:bulk_ph}.

\newpage
\subsection*{Supplementary Table \ref{tab:force_bop_compare}: Accuracy Validation}

\begin{table}[htbp]
    \centering
    \begin{tabular}{|c|c|c|c|c|c|c|}   
        \hline
       	 & \multicolumn{2}{c|}{Crystal}  & 
        \multicolumn{2}{c|}{Transition} & \multicolumn{2}{c|}{Molten} \\
        \hline
         DFT & I & II & I & II & I & II\\
        \hline
        BOP & 0.254 & 0.241 & 0.322 & 0.376 & 0.258 & 0.305 \\ 
        \hline
        MGP & 0.082 & 0.120 & 0.148 & 0.206 & 0.110 & 0.139 \\ 
        \hline
    \end{tabular}
    \caption{Mean absolute error (MAE) of the force prediction compared to DFT, in units of \eVperA.}
    \label{tab:force_bop_compare}
\end{table}

We compare the performance of the MGP against the BOP force field of Ref. \cite{cherukara2016ab}. 
In particular, we validate results using frames of different atomic structures taken from an independent MD trajectory simulating a melting process of stanene at temperature $T\approx 500$K, which contains 2D crystal lattice, crystal-to-liquid transition state and disordered liquid configurations. 
Then we compute the mean absolute error of forces with respect to the DFT prediction.
Since the MGP force field is trained with PBEsol exchange-correlation functional, ultra-soft pseudopotential, of the platform Quantum Espresso (labelled as DFT ``I''), while the BOP potential is trained with PBE functional, PAW pseudopotential of VASP (labelled as DFT ``II''), we make a comparison of the mean absolute error of force prediction with both DFT ``I'' and ``II'' as the ground truth.
In Table \ref{tab:force_bop_compare} we show that the MGP achieves a better accuracy than the BOP in all of these frames, by at least 50\%. As temperature is increased, and as the deviation from the crystal phase increases, both force fields have larger MAE. This is because the system at high temperature explores many more different configurations, likely different from the training set, nevertheless, the results show that the MGP has uniformly better accuracy.

\newpage
\section*{Supplementary Figures}

\begin{figure}[htbp]
    \centering
    \includegraphics{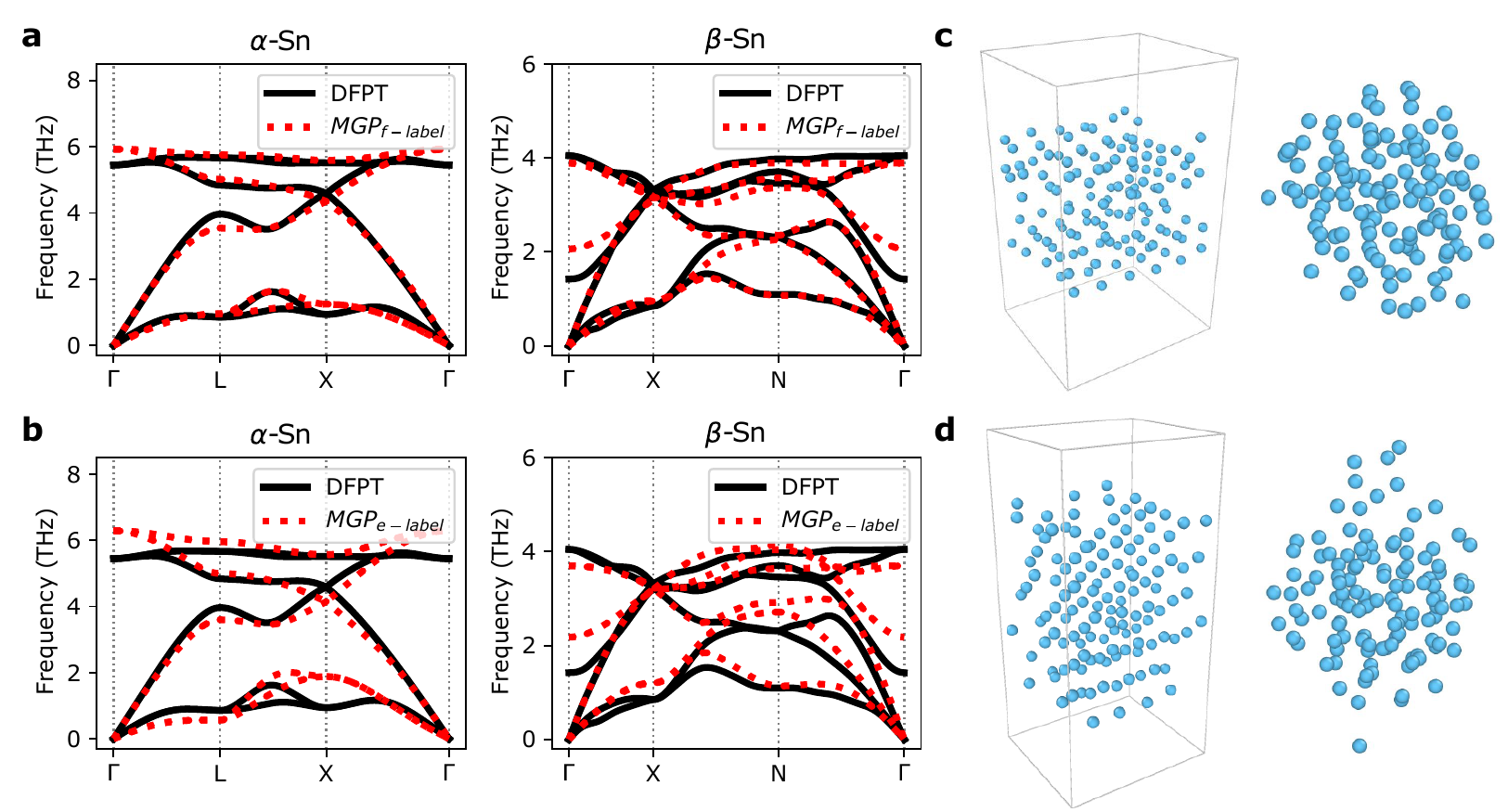}
    \caption{
    \textbf{a.} The phonon dispersions of $\alpha$-Sn and $\beta$-Sn predicted by $\text{MGP}_{f-label}$. 
    \textbf{b.} The phonon dispersions predicted by $\text{MGP}_{e-label}$.
    \textbf{c.} The resulting configurations of monolayer-bulk (left) and monolayer-liquid (right) transitions from $\text{MGP}_{f-label}$. 
    \textbf{d.} The resulting configurations of monolayer-bulk (left) and monolayer-liquid (right) transitions from $\text{MGP}_{e-label}$.  }
    \label{supp-fig:bulk_ph}
\end{figure}

\subsection*{Supplementary Figure \ref{supp-fig:bulk_ph}: Bulk Polymorphism}

We compute the phonon dispersions from $\text{MGP}_{f-label}$ and $\text{MGP}_{e-label}$ presented in Supplementary Table \ref{supp-tab:bulk_properties}. The results compared with the density functional perturbation theory (DFPT) are shown in Fig.\ref{supp-fig:bulk_ph}(a)(b).
The monolayer-bulk and monolayer-liquid transitions of stanene obtained with the initial BFF are qualitatively similar to those from both updated force fields with bulk phase training data included, as shown in Fig.\ref{supp-fig:bulk_ph}(c)(d), with ordered bulk and liquid phases emerging at the end of transitions.
We note that the model without bulk phase training data has low force errors of 0.064 and 0.063 \eVperA for $\alpha$-Sn and $\beta$-Sn, while $\text{MGP}_{f-label}$ and $\text{MGP}_{e-label}$ that include bulk phase training data further improves the BFF accuracy for these bulk phases, as expected. 
$\text{MGP}_{f-label}$ presents a good match of phonon spectrum while the predicted bulk modulus is not satisfactory. Adding global energies as additional training labels resolves this issue and renders accurate lattice constant and bulk modulus, as shown by $\text{MGP}_{e-label}$, while the phonon spectrum becomes somewhat less accurate due to the slight degradation of force accuracy.
This is an indication that the GP BFF performance is determined by the choice of training data labels, likely due to the limited descriptive power of the 2+3 body interaction model. 
We note that, given the computational speed approaching classical force fields, our model can describe well the forces and hence dynamics in different phases, but careful selection of the training labels is needed if high accuracy is desired in multiple properties related to absolute energies and their second derivatives. A separate effort is underway focused on developing kernels with higher descriptive power to expand the capability of MGP models to capture multiple structural and dynamical properties.

\begin{figure}[htbp]
    \centering
    \includegraphics[width=\textwidth]{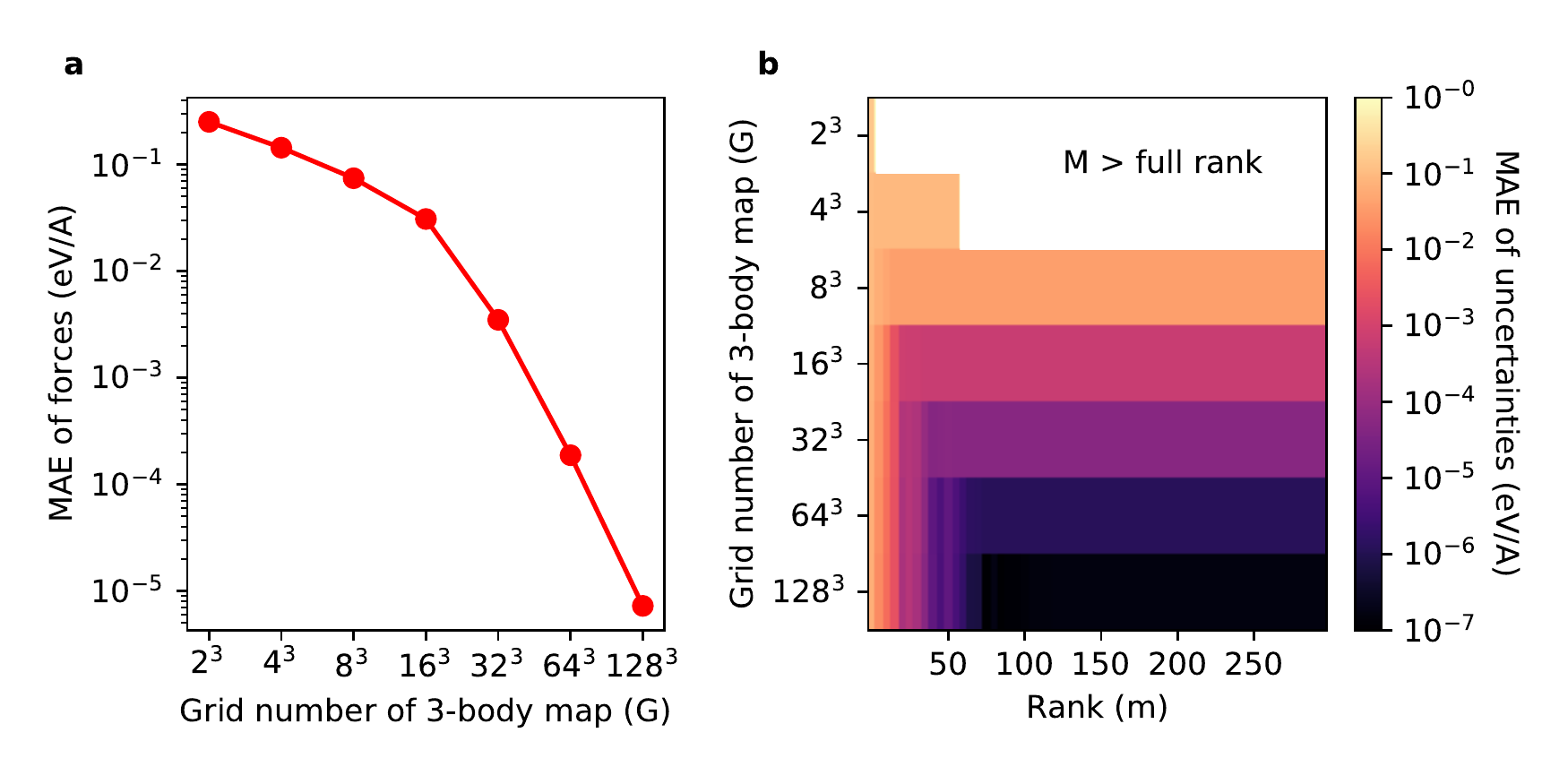}
    \caption{
    \textbf{a.} Convergence of the MGP force w.r.t. the GP forces with different grid numbers ($G$). 
    \textbf{b.} Convergence of the MGP uncertainty prediction w.r.t. the GP uncertainty with different grid numbers ($G$) and SVD ranks ($m$)}
    \label{supp-fig:convergence}
\end{figure}

\newpage
\subsection*{Supplementary Figure \ref{supp-fig:convergence}: Convergence}

As a demonstration of the accuracy of the mapping method, a training set of 100 atomic environments from the stanene system is used to build up the Gaussian Process model, based on which the MGP with mapped force and uncertainty is constructed. Then the GP and MGP are used to predict both the forces and uncertainties for a testing configuration, and the mean absolute error (MAE) of the MGP predictions with respect to the GP predictions are computed.

The Gaussian process model uses 2+3-body kernel functions with optimized hyperparameters. The 2-body component and the 3-body component are mapped separately and summed over in prediction. 
The uncertainty mapping for the 2-body component is cheap in terms of both the computational cost and storage requirement, and is fixed using 144 grid points and full rank 144. 
The uncertainty mapping for the 3-body component is constructed with different grid numbers, ranging from $2 \times 2 \times 2$ to $128 \times 128 \times 128$, as presented in the y-axis of Fig.\ref{supp-fig:convergence}b. For each grid number, different ranks are used for prediction, ranging from $5$ to $3\Ntrain = 300$ (the full rank with $\Ntrain =100$ training points), as presented in the x-axis of Fig.\ref{supp-fig:convergence}b. 
Smaller mean absolute error (MAE) between MGP uncertainties and GP uncertainties is indicated by darker color in Fig.\ref{supp-fig:convergence}b, while for small grid numbers $2\times 2\times 2$ and $4\times 4\times 4$ the full rank is small, thus there are no testing data from higher ranks, as shown in the slash line area.

From the convergence plot Fig.\ref{supp-fig:convergence}, we observe that the mapping with a relatively small grid number of $32 \times 32 \times 32$ and only $\sim 10\%$ full rank can reach a mean absolute error below the level of 1 m\eVperA. Larger number of grid points increases the accuracy since the interpolation mesh becomes finer. In addition, given a grid number the accuracy saturates after the rank exceeds a value much smaller than the full rank, which implies the information in the kernel matrix can be extracted with a low-rank approximation, therefore, justifies our mapping approach of the uncertainty prediction.

\begin{figure}[htbp]
    \centering    
    \includegraphics[width=0.75\textwidth]{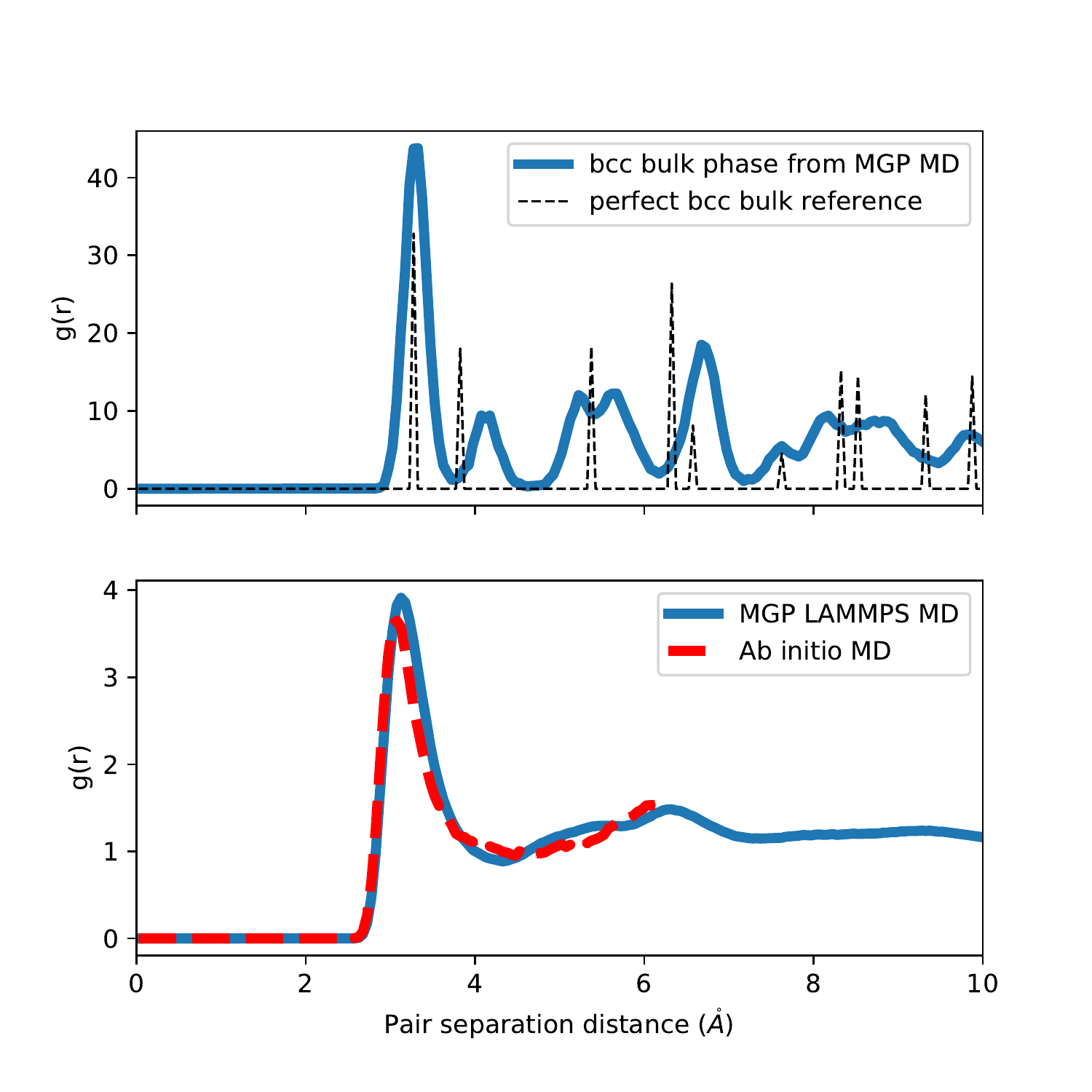}
    \caption{Radial distribution function (RDF) of 3D slab and liquid}
    \label{supp-fig:rdf}
\end{figure}

\newpage
\subsection*{Supplementary Figure \ref{supp-fig:rdf}: Phase Transition Process}

The radial distribution function (RDF) of the slab configuration from 200K MD simulation is shown in Fig.\ref{supp-fig:rdf} (top), corresponding to the configuration at 2ns in Fig.4(a) of the main text. The black dash line is the RDF of the perfect bcc lattice of tin \cite{mp_7162_sn}. As shown in Fig.4(a)-2ns, besides the orange atoms which are identified as bcc sites by the common neighbor analysis, the atoms near the surfaces and grain boundaries are not recognized as any known crystal lattice sites. Therefore, the RDF of this configuration does not match with the bcc phase perfectly. 

The radial distribution function of 600 ps frame (the liquid phase configuration) in the melting process simulation is shown in Fig.\ref{supp-fig:rdf} (bottom), which shows a peak at $\sim 3.1\ \text{\AA}$ and a tail that converges to a uniform density. 
To validate this liquid phase, we run an ab-initio molecular dynamics simulation of a small system (64 atoms) in the liquid phase for 10 ps, and compute the radial distribution function from the trajectory. 
The AIMD cell size is 12.22 \AA\ in each dimension, so the RDF is truncated at 6.1 \AA.
The result is shown as the red dash line, which aligns well with our MD simulation with the MGP force field. 
This indicates that the final cluster becomes a 3D disordered dense liquid phase of Sn.

The videos of the phase transformation process from MD simulation can be downloaded from 
\href{https://doi.org/10.24435/materialscloud:qg-99}{DOI: 10.24435/materialscloud:qg-99} \cite{data}.

\begin{figure}[htbp]
    \centering
    \includegraphics{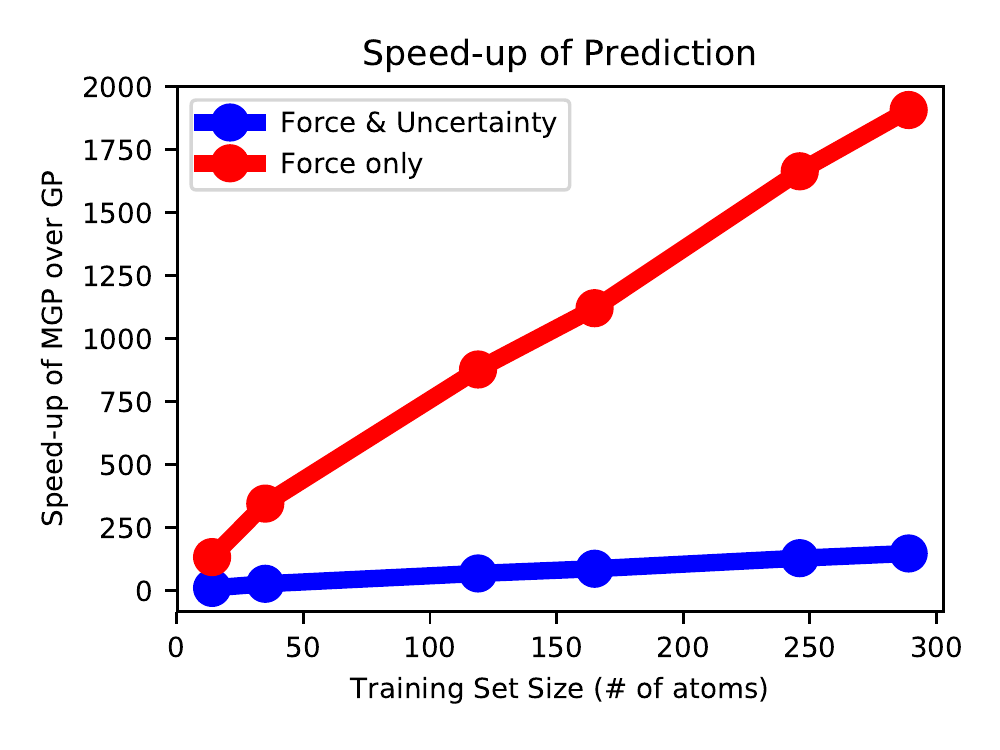}
    \caption{Speed-up of MGP compared to GP grows linearly w.r.t. the training size. Prediction of both force and uncertainty reaches 2 orders of magnitude of acceleration when GP has 300 training data, while force-only prediction is 3 orders of magnitude faster.}
    \label{supp-fig:mgp-speedup}
\end{figure}

\newpage
\subsection*{Supplementary Figure \ref{supp-fig:mgp-speedup}: Performance}

Since the original GP is a non-parametric method, it depends on the training data set completely, thus its prediction cost scales linearly with respect to the training data set size. After the force field is trained and mapped, MGP is obtained as a parametric model, whose basis is a group of cubic spline functions, and can be directly employed independent of the original GP regression model as well as training data set, thus reaches constant scaling of computational cost at prediction. 
The major computational bottleneck in the prediction of the current scheme lies in two parts. One is the construction of the local atomic environment, which requires calculating interatomic distances and identifying all $n$-atom subsets. 
The other is the calculation of the first term in the variance formulation, the self-kernel function $k_{\alpha\alpha}(\env, \env)$ which scales quadratically with the number of $n$-atom subsets in $\env$, while the second term, the vectorized spline function $\hat{\boldsymbol{\phi}}$, scales linearly. 
We note the situation is different in the original GP, where the construction of atomic environment and the evaluation of the self-kernel are negligible compared to the evaluation of the kernel vector. In this way the present mapping scheme for both forces and uncertainties eliminated the main computational bottleneck of the full GP model.

\newpage
\bibliographystyle{naturemag}
\bibliography{supplementary}